\documentstyle[12pt]{article}

\newcommand{\postscript}[2]{\vspace 5cm}
\newcommand{\postbb}[3]{\vspace 5cm}

\font\tenbf=cmbx10
\font\tenrm=cmr10
\font\tenit=cmti10
\font\elevenbf=cmbx10 scaled\magstep 1
\font\elevenrm=cmr10 scaled\magstep 1
\font\elevenit=cmti10 scaled\magstep 1

\evensidemargin
0.30truein
\renewenvironment{thebibliography}[1]
 { \elevenrm
   \begin{list}{\arabic{enumi}.}
    {\usecounter{enumi} \setlength{\parsep}{0pt}
     \setlength{\itemsep}{3pt} \settowidth{\labelwidth}{#1.}
     \sloppy
    }}{\end{list}}

\def\Tr{{\rm Tr}}
\def\ppm{\mbox{$\pm$}}
\newcommand{\beqa}{\begin{eqnarray}}
\newcommand{\eeqa}{\end{eqnarray}}
\def\lag{Lagrangian}
\def\ra{\rightarrow}
\newcommand{\rmm}[1]{{\rm #1}}
\newcommand{\shz}{\mbox{$\hat{s}^2_Z$}}

\newcommand{\RMP}[3]{{\elevenit Rev. Mod. Phys.} {\elevenbf  #1}, #2 (19#3)}
\newcommand{\PR}[3]{{\elevenit Phys. Rev.} {\elevenbf #1}, #2 (19#3)}
\newcommand{\PL}[3]{{\elevenit Phys. Lett.} {\elevenbf #1}, #2 (19#3)}
\newcommand{\Rep}[3]{{\elevenit Phys. Rep.} {\elevenbf #1}, #2 (19#3)}

\newcommand{\PRL}[3]{{\elevenit Phys. Rev. Lett.} {\elevenbf #1}, #2 (19#3)}
\newcommand{\NP}[3]{{\elevenit Nucl. Phys.} {\elevenbf #1}, #2 (19#3)}
\newcommand{\ZP}[3]{{\elevenit Zeit. Phys.} {\elevenbf #1}, #2 (19#3)}
\newcommand{\con}[3]{{\elevenbf #1}, #2 (19#3)}

\newcommand{\msb}{\mbox{$\overline{\rm{MS}}\ $}}
\newcommand{\mt}{\mbox{$m_t$}}

\newcommand{\mz}{\mbox{$M_Z$}}

\newcommand{\alsz}{\mbox{$\alpha_s(M_Z)$}}
\newcommand{\als}{\mbox{$\alpha_s$}}

\newcommand{\stacksub}[2]{\ _{\stackrel{\textstyle
#1}{\scriptstyle #2}}\ }
\newcommand{\skipblk}[1]{}
\def\bqa{\begin{eqnarray}}
\def\eqa{\end{eqnarray}}

\newcommand{\doub}[3]{\mbox{$
\left( \begin{array}{c} #1    \\ #2  \end{array} \right)_#3$}}

\newcommand{\etal}{{\elevenit et al., }}

\newcommand{\eg}{{\elevenit e.g., }}
\newcommand{\ie}{{\elevenit i.e., }}
\newcommand{\stto}{\mbox{$SU_{2L} \x SU_{2R} \x U_1\ $}\ }
\newcommand{\sto}{\mbox{$SU_2 \x U_1\ $}}

\newcommand{\x}{\mbox{$\times$}}

\newcommand{\sinn}{\mbox{$\sin^2\theta_W\,$}}

\newcommand{\beq}{\begin{equation}}
\newcommand{\eeq}{\end{equation}}

\newcommand{\RA}{\mbox{$\rightarrow$}}

\def\mxth{\mathsurround=0pt }
\def\xversim#1#2{\lower2.pt\vbox{\baselineskip0pt \lineskip-.5pt
  \ialign{$\mxth#1\hfil##\hfil$\crcr#2\crcr\sim\crcr}}}
\def\simgr{\mathrel{\mathpalette\xversim >}}

\begin{document}
\begin{center}{{\tenbf TESTS OF THE STANDARD MODEL AND
   SEARCHES FOR NEW PHYSICS\footnote{To be published in
   {\it Precision Tests of the Standard Electroweak Model}, ed.
   P. Langacker (World, Singapore, 1994).}\\}
\vglue 1.0cm
{\tenrm PAUL LANGACKER \\}
\baselineskip=13pt
{\tenit Department of Physics, University of Pennsylvania,\\
 Philadelphia, Pennsylvania, USA 19104-6396 \\}}
\vglue 0.8cm
\end{center}

\vglue 0.6cm
\tableofcontents

\baselineskip=14pt
\elevenrm

\section{\elevenbf Introduction}
\vglue 0.4cm
Despite its many successes, the standard model cannot be taken
seriously as a candidate for the ultimate theory of matter.
As described in the article {\it Structure of the Standard
Electroweak Model} earlier in this volume,
it is a complicated theory with
many free parameters, several fine-tuning problems,
and many arbitrary and unexplained features.
Historically, precision electroweak experiments were crucial
for establishing the standard electroweak model (especially the
unification aspects) as correct to first approximation and
excluding alternatives. At present and in the future they
will continue to be important for establishing the domain of
validity of the standard model, determining its parameters,
and searching for and excluding various possibilities for
underlying new physics.

Earlier chapters of this volume have described the standard
model and its renormalization, the various types of precision
experiments, and their implications in detail. This chapter
is devoted to global analyses of the $Z$-pole, $M_W$, and
neutral current data\footnote{In most cases, the types
of new physics constrained by charged current data (Chapter V)
are separate and do not require a simultaneous analysis with
the gauge boson and neutral current data. One major exception is
mixing between ordinary and exotic fermions.
See the article by D. London in this volume.},
which contains more information than any one
class of experiment. The subsequent sections will summarize
some of the relevant data and theoretical formulas, the status of
the standard model tests and parameter determinations,
the possible classes of new physics,
and the implications of the precision experiments.
In particular, the model independent analysis of neutral current couplings
(which establishes the standard model to first approximation);
the implications of supersymmetry; (supersymmetric)
grand unification; and a number
of specific types of new physics, including heavy $Z^\prime$
bosons\footnote{\stto models
(which involve an additional heavy $W^\prime$ boson coupling
to right-handed currents) are described in the chapters by
Deutsch and Quin and by
Herczeg, and in \cite{lrmodels}.} ,
new souces of $SU_2$ breaking, new contributions to the
gauge boson self-energies, $Zb\bar{b}$ vertex corrections,
certain types of new 4-Fermi operators and leptoquarks, and
exotic fermions are described. Leptoquarks are covered
in  much more detail in the chapters by Deutsch and Quin and by
Herczeg, and exotic fermions in the chapter by London.
Future prospects are described in the chapter by Luo and in
\cite{LLM}.

\vglue 0.6cm
\section{\elevenbf The Standard Model and its Parameters}

\subsection{\elevenit Recent Data}
\vglue 0.4cm

Recent results from $Z$-pole experiments are shown in Table~\ref{tab1}.
These include the results of the four LEP experiments ALEPH,
DELPHI, L3, and OPAL (including
preliminary results from the 1993 LEP energy scan), averaged including
a proper treatment of common systematic uncertainties~\cite{a1,a1a}.  In
addition, the result from the SLD experiment at SLAC~\cite{a2} on the
left-right asymmetry $A_{LR}$ is shown.  The first row in Table~\ref{tab1}
gives the value of the $Z$ mass, which is now known to remarkable
precision.  Also shown are the lineshape variables $\Gamma_Z$, $R$, and
$\sigma_{\rm had}$, which are respectively the total $Z$ width,
the ratio of the $Z$ width into hadrons to
the width into a single charged lepton,
and the peak hadronic cross section after removing QED effects; the heavy quark
production rates; various
forward-backward asymmetries, $A_{FB}$; quantities derived from the $\tau$
polarization $P_{\tau}$ and its angular distribution; and the effective
weak angle $\bar{s}^2_\ell$ obtained from the jet charge asymmetry. $N_\nu$
is the number of effective active neutrino flavors with masses light enough
to be produced in $Z$ decays.  It is obtained by subtracting the widths for
decays into hadrons and charged leptons from the total width $\Gamma_Z$
obtained from the lineshape.  The asymmetries are expressed in terms of the
quantity
\beq A^o_f = \frac{2 \bar{g}_{Vf} \; \bar{g}_{Af}}{\bar{g}^2_{Vf}
+ \bar{g}^2_{Af}},  \label{eqn1} \eeq
where $\bar{g}_{V,Af}$ are the vector and axial vector couplings to fermion
$f$.

\begin{table} \centering \small
\begin{tabular}{|l|c|c|}
\hline \hline
Quantity & Value & Standard Model \\ \hline
$M_Z$ (GeV) & $91.1888 \pm 0.0044$ &  input \\
$\Gamma_Z$ (GeV) & $2.4974 \pm 0.0038$ & $2.497 \pm 0.001 \pm
0.003 \pm [0.002] $ \\
$R = \Gamma({\rm had})/\Gamma(\ell \bar{\ell})$ & $20.795 \pm
0.040$ & $20.784 \pm 0.006 \pm 0.003 \pm [0.03]$ \\
$\sigma_{\rm had} = \frac{12 \pi}{M_Z^2} \; \frac{\Gamma(e
\bar{e}) \Gamma({\rm had})}{\Gamma_Z^2} ({\rm nb})$ & $41.49 \pm 0.12$
& $41.44 \pm 0.004 \pm 0.01 \pm [0.02]$ \\
$R_b = \Gamma(b \bar{b})/ \Gamma({\rm had})$ &$0.2202 \pm 0.0020$
& $0.2156 \pm 0 \pm 0.0004$ \\
$R_c = \Gamma(c\bar{c}) /\Gamma({\rm had})$ & $0.1583 \pm 0.0098$
 & $0.171 \pm 0 \pm 0$ \\
$A^{0\ell}_{FB} = \frac{3}{4} \left( A_{\ell}^0 \right)^2$ &
$0.0170 \pm 0.0016$ & $0.0151 \pm 0.0005 \pm 0.0006$  \\
$A^0_{\tau} \left(P_\tau \right)$ & $0.143 \pm 0.010$ & $0.142
\pm 0.003 \pm 0.003$ \\
$A^0_e \left( P_\tau\right)$ & $0.135 \pm 0.011$ & $0.142 \pm
0.003 \pm 0.003$ \\
$A^{0b}_{FB} = \frac{3}{4} A^0_e A^0_b$ & $0.0967 \pm 0.0038$ &
$0.0994 \pm 0.002 \pm 0.002$ \\
$A^{0c}_{FB} = \frac{3}{4} A^0_e A^0_c$ & $0.0760 \pm 0.0091$ &
$0.071 \pm 0.001 \pm 0.001$ \\
$\bar{s}^2_{\ell} \left(A^{Q}_{FB} \right)$ & $0.2320 \pm
0.0016$ & $0.2322 \pm 0.0003 \pm 0.0004$ \\
$A^0_e \left(A^0_{LR} \right)$ \ \ (SLD) & $0.1637 \pm 0.0075
\;\; (92 + 93)$ & $0.142 \pm 0.003 \pm 0.003$ \\
     & $(0.1656 \pm 0.0076 \;\; (93))$ & \ \\
$N_\nu$ & $2.988 \pm 0.023$ & $3$ \\ \hline
\end{tabular}
\caption[]{$Z$-pole observables from LEP and SLD compared to their standard
model expectations.  The standard model prediction is based on $M_Z$ and
uses the global best fit values for $m_t$ and $\alpha_s$, with $M_H$ in the
range $60 - 1000$~GeV. The $R_b-R_c$ correlation is $-0.4$. The lineshape
correlations are given in \cite{a1a}.}
\label{tab1}
\end{table}

{}From the $Z$ mass one can predict the other observables,
including electroweak loop effects.  The predictions also depend on the top
quark and Higgs mass, and $\alpha_s$ is needed for the QCD corrections to
the hadronic widths and the relation between $M_Z$ and \sinn.
The predictions are shown in the third column of Table~1,
using the value $m_t = 175 \pm 11 $~GeV obtained for $M_H = 300$~GeV in
a global best fit to all data.  The first uncertainty is from $M_Z$ and
$\Delta r$ (related to the running of $\alpha$ up to $M_Z$), while the
second is from $m_t$ and $M_H$, allowing the Higgs mass to vary in the
range $60 - 1000$~GeV.  The last uncertainty is the QCD uncertainty from
the value of $\alpha_s$.  Here the value and uncertainty are given by
$\alpha_s = 0.127 \pm 0.005$, obtained from the global fit to the lineshape.

The data is in excellent agreement with the standard model predictions
except for two observables.  The first is
\beq R_b = \frac{\Gamma(b\bar{b})}{\Gamma(\rm had)} = 0.2202 \pm
0.0020.\eeq
This is some $2.3 \sigma$ higher than the standard model expectation
$0.2156 \pm 0.0004$.  Because of special vertex corrections, the $b
\bar{b}$ width actually decreases with $m_t$, as opposed to the other
widths which all increase.
It is apparent from Figure~\ref{fig1}
that $R_b$ favors a small value of $m_t$.  By itself $R_b$
is insensitive to $M_H$.  However, when combined with other observables,
for which $m_t$ and $M_H$ are strongly correlated, the effect is to favor a
smaller Higgs mass.  Another possibility, if the effect is more than a
statistical fluctuation, is that it may be due to some sort of new physics.
Many types of new physics will couple preferentially to the third
generation, so this is a serious possibility.

\begin{figure}
\postbb{40 220 530 680}{/home/pgl/fort/nc/graph/gam/xxrb.ps}{0.6}
\caption[]{ Standard model prediction for $R_b \equiv \Gamma(b \bar{b}) /
\Gamma( \rm had)$ as a function of $m_t$, compared with the LEP
experimental value.  Also shown are the D0 lower bound of 131~GeV and the
CDF range $174 \pm 16$~GeV.}
\label{fig1}
\end{figure}

Despite the small discrepancy, $R_b$ and the forward-backward asymmetry
$A^{0b}_{FB}$ (corrected for $B \bar{B}$ oscillations)
are sufficient to establish that the left-handed $b$
belongs to a weak doublet. From the
fit to these and other data one obtains uniquely
\beq  t_{3L}(b) = -0.500 \pm 0.005 \ \ \ \ \ \ \
t_{3R}(b) = 0.026 \pm 0.018 \label{biso} \eeq
for the third component of the weak isospin of the $b_{L,R}$, respectively,
updating an analysis by Schaile and Zerwas \cite{v16}. This is
in agreement with the standard model expectations of $-1/2$ and $0$ and
excludes topless models\footnote{Earlier indirect arguments
for the existence of the $t$ quark are summarized in~\cite{v15}.}.
The values of $R_b$ and
$A^{0b}_{FB}$ are also compared with the predictions of various
alternative \sto models in  Table~\ref{b}. Only the standard model
is in agreement with the data.

\begin{table}                   \centering
\small \def\baselinestretch{1} \normalsize
\begin{tabular}{|c|c|c|c|c|c|} \hline
Quantity & Experiment & SM & Topless & Mirror & Vector \\
\hline
$R_b$ &
0.2202 \ppm 0.0020 & 0.2156 & 0.017 & 0.2156 & 0.35 \\
$A^{0b}_{FB} $ & 0.0967 \ppm 0.0038 & 0.0994 & 0 & $-0.0994$ & 0 \\ \hline
\end{tabular}
\caption[]{Predictions of the standard model (SM),
topless models, a mirror
model with $(t \; b)_R$ in a doublet, and a vector model with left and
right-handed doublets, for $R_b$
and $A^{0b}_{FB}$, compared with the experimental values.}
\label{b}
\end{table}

The other discrepancy is the value of the left-right asymmetry
\beq A^0_{LR} = A^0_e = \frac{2 \bar{g}_{Ve}
\bar{g}_{Ae}}{\bar{g}_{Ve}^2 + \bar{g}_{Ae}^2} = 0.164 \pm 0.008
\eeq
obtained by the SLD collaboration.
$A^0_{LR}$ is very clean both experimentally \cite{blond}
(most systematic effects other than the absolute beam polarization
cancel in the ratio) and theoretically (most radiative corrections
cancel), and is very sensitive to both the weak angle and new physics.
The SLD value is some $2.5 \sigma$ higher than
the standard model expectation of $0.142 \pm 0.004$.
Assuming the standard model, $A^0_{LR}$ (combined with $M_Z$) predicts
a top quark mass around 250 GeV, much higher than other determinations.
Unless this is a statistical fluctuation, the obvious possibility
is that the high value of $A^0_{LR}$ is due to new physics,
such as $S < 0$,
where $S$ is a parameter describing certain types of heavy new physics
(see Section \ref{stusec}).  In
addition, there are possible tree-level physics such as heavy $Z'$ bosons
or mixing with heavy exotic doublet leptons, $E'_R$, which could
significantly affect the asymmetry.  However, new physics probably cannot
for the entire discrepancy, because
the LEP observables
$A^{0e}_{FB}$ and $A^0_e \left( P_\tau\right)$ (obtained
from the angular distribution of the $\tau$ polarization)
measure the same quantity\footnote{The
relation makes use only of the
assumption that the LEP and SLD observables are dominated by the $Z$-pole.
The one (unlikely) loophole is the possibility of an important
contribution from other sources, such as new 4-fermi operators.  These are
mainly significant slightly away from the pole (at the pole they are out of
phase with the $Z$ amplitude and do not interfere).}
$A^0_e$ as $A^0_{LR}$.
Together, the LEP values imply
\beq A^0_e \left|_{\rm LEP} = 0.138 \pm 0.009 \right. , \eeq
in agreement with the standard model expectation.
Thus, there is a direct experimental conflict between the LEP and SLD
values of $A_e^0$ at the $2.2\sigma$ level.

It will take more time and more statistics to see whether the values of
$R_b$ and $A^0_{LR}$ constitute true discrepancies with the standard
model (and between experiments in the latter case) or are due
to statistical fluctuations at the 2-2.5 $\sigma$ level or other
experimental problems. I will generally take the view that  the
effects are consistent with (large) fluctuations.

There are many other precision  observables.  Some recent ones
are shown in Table~\ref{tab2}.  These include the D0 limit \cite{a2a} $m_t
> 131$~GeV and the value $m_t = 174 \pm 16$~GeV suggested by the CDF
candidate events \cite{a2b}.  There are new observations of the $W$ mass
\cite{a3i,a3} from both D0, which has presented a preliminary new value $79.86
\pm 0.40$~GeV, and from CDF, which finds $80.38 \pm 0.23$~GeV.  Combining
these and earlier data one obtains the results shown.  Other
observables include $M_W/M_Z$ from UA2\footnote{One could, of course,
multiply $M_W/M_Z$ by the LEP $M_Z$ and include the result in the $M_W$
average.  (In fact, such a procedure was carried out in the D0 analysis.)
I do not do so because, in principle, it would introduce a correlation
between $M_Z$ and $M_W$.  In practice, the effect is negligible because of
the tiny uncertainty in $M_Z$.}~\cite{a4},
atomic parity violation from Boulder~\cite{v4},
recent results on neutrino
electron scattering from CHARM II~\cite{a5}, and new measurements of $s_W^2
\equiv 1 - M^2_W/M^2_Z$ from the CCFR collaboration at Fermilab~\cite{a6}.
This on-shell definition of the weak angle is determined from deep
inelastic neutrino scattering with small sensitivity\footnote{The
sensitivity is small but not zero. The quoted CCFR value is for
$(m_t,M_H)$ = (150,100) GeV. The combined result is for
$(m_t,M_H)$ in the allowed range.}
to the top quark mass.
The result combined with earlier experiments~\cite{a7i}-\cite{a7a}
is also shown.  All
of these quantities are in excellent agreement with the standard model
predictions.
\begin{table} \centering \small
\begin{tabular}{|c|c|c|}  \hline \hline
Quantity & Value & Standard Model \\ \hline
$M_W$ (GeV) & $80.17 \pm 0.18$ & $80.31 \pm 0.02 \pm 0.07$ \\
$M_W/M_Z (UA2)$ & $0.8813 \pm 0.0041$ & $0.8807 \pm 0.0002 \pm
0.0007$ \\
$Q_W (C_S)$ & $-71.04 \pm 1.58 \pm [0.88]$ & $-72.93 \pm 0.07 \pm
0.04$ \\
$g_A^{\nu e}$ (CHARM II) & $-0.503 \pm 0.017$ & $-0.506 \pm 0 \pm
0.001$ \\
$g_V^{\nu e}$ (CHARM II) & $-0.035 \pm 0.017$ & $-0.037 \pm 0.001
\pm 0$ \\
$s^2_W \equiv 1 - \frac{M_W^2}{M_Z^2}$ & $\begin{array}{c} 0.2218
\pm 0.0059 \;{\rm [CCFR]} \\ 0.2260 \pm 0.0048 \; {\rm [All]}
\end{array}$ & $0.2245 \pm 0.0003\pm 0.0013$ \\
$M_H$ (GeV) & $\geq 60$ LEP & $< \left\{ \begin{array}{c} 0
(600), \; {\rm theory} \\ 0 (800), \; {\rm indirect} \end{array}
\right.$ \\
$m_t$ & $> 131$ D0 & $175 \pm 11 ^{+17}_{-19}$ [indirect]
\\
 \ & $174 \pm 16$ CDF & \ \\
$\alpha_s (M_Z)$ & $\begin{array}{cc} 0.123 \pm 0.006 & {\rm LEP
\; jets} \\ 0.116 \pm 0.005 & {\rm jets
}+{\rm
low \; energy} \end{array}$ & $\begin{array}{c} 0.127 \pm 0.005
\pm 0.002 \\ \left[ Z \; \; {\rm lineshape}\right] \end{array}$
\\ \hline
\end{tabular}
\caption[]{Recent observables from the $W$ mass and other
non-$Z$-pole observations compared with the standard model
expectations.  Direct values and limits on $M_H$, $m_t$, and
$\alpha_s$ are also shown.}
\label{tab2}
\end{table}

In the global fits to be described~\cite{a7c}, all of the earlier low energy
observables~\cite{a7}-\cite{a7b} not listed in the table are fully
incorporated,
as are full treatments of statistical, systematic, and theoretical
uncertainties, and correlations between the experiments.

\vglue 0.6cm
\subsection{\elevenit Theoretical Expressions and Radiative Corrections}
\subsubsection{\elevenit The $Z$ and $W$ Masses}
\vglue 0.4cm

In the electroweak theory one defines the weak angle by
\beq \sin^2 \theta_W \equiv \frac{g'^2}{g^2 + g'^2}
\longrightarrow \sin^2 \hat{\theta}_W (M_Z) \;\;\;\;
(\overline{MS})   \label{eq1}\eeq
where $g'$ and $g$ are respectively the gauge couplings of the $U_1$ and
$SU_2$ gauge groups.  Although initially defined in terms of the gauge
couplings, after spontaneous symmetry breaking one can relate the weak angle
to the $W$ and
$Z$ masses  by
\beq M_W^2 = \frac{A^2}{\sin^2 \theta_W} \longrightarrow
\frac{A^2}{\sin^2 \hat{\theta}_W ( 1- \Delta \hat{r}_W)}
\label{eq2} \eeq and
\beq M_Z^2 = \frac{M^2_W}{\cos^2 \theta_W} \longrightarrow
\frac{M^2_W}{\hat{\rho} \cos^2 \hat{\theta}_W} \label{eq3} \eeq
where
\beq A^2 \equiv \frac{\pi \alpha}{\sqrt{2} G_F} = \left( 37.2802 \
{\rm GeV} \right)^2. \label{adef} \eeq
The first form of equations (\ref{eq1})--(\ref{eq3}) are valid at tree
level.  However, the data is sufficiently precise that one must include
full one loop radiative corrections, which means that one must replace the
quantities by the expressions shown in the last part of equations.  There
are a number of possible ways of defining the renormalized weak angle.
Here I am using the quantity $\sin^2 \hat{\theta}_W(M_Z) \equiv \shz$,
which is renormalized according to modified minimal subtraction,
$\overline{MS}$~\cite{a8}.  This basically means that one removes the
$\frac{1}{n-4}$ poles and some associated constants (artifacts
of dimensional regularization) from the gauge couplings.

In equation (\ref{eq2}) the quantity $\Delta \hat{r}_W$ contains
the finite radiative corrections which relate the $W$ and $Z$ masses, muon
decay, and QED.  The dominant contribution is given by the running of the
fine structure constant $\alpha$ from low energies, where it is defined in
QED, up to the $Z$-pole, which is the scale relevant for electroweak
interactions,
\beq \frac{1}{1 - \Delta \hat{r}_W} \simeq \frac{\alpha (M_Z)}{\alpha} \sim
\frac{1/128}{1/137}. \eeq
There is only a weak dependence on the top quark mass in this scheme,
leading to a value \beq \Delta \hat{r}_W \sim 0.07 \label{eqa6} \eeq
dominated by the running of $\alpha$.
More precisely, in the \msb scheme~\cite{a13}
\beq \alpha^{-1}(M_Z) = 127.9 \pm 0.1,  \eeq
where the uncertainty is
from the contribution of light hadrons to the photon self-energy diagrams.
This leads to a theoretical uncertainty of $\pm 0.0009$~\cite{a8a}
in $\Delta \hat{r}_W$, which turns out to
be the dominant theoretical uncertainty in the precision electroweak tests
and, in particular, in the expressions relating the $Z$ mass to other
observables.  A similar effect leads to a significant theoretical
uncertainty in  the anomalous magnetic moment of the muon,
$g_\mu - 2$, which will dominate the experimental
uncertainties in the new Brookhaven experiment unless associated
measurements are made of the cross-section for $e^+e^- \ra $~hadrons at low
energies. Including other small contributions, one predicts
\beq \Delta \hat{r}_W = 0.0706 \pm 0.0002 \pm 0.0009, \label{delrhat} \eeq
where the central value is for $(m_t,M_H)$ = (175, 300) GeV, the
first uncertainty is from $m_t$ and $M_H$ varying in the global
best fit range, and the second is from $\alpha(M_Z)$.

Because $m_t$ is so much heavier than the bottom quark mass there
is large $SU_2$ breaking generated by loop diagrams involving the
top and bottom quarks, in particular from the $W$ and $Z$
self-energy diagrams shown in Figure~\ref{fig1a}.
\begin{figure}
\postbb{60 200 550 600}{/home/pgl/fort/nc/graph/misc/self.ps}{0.5}
\caption[]{(a) Photon self-energy diagram leading to the running of
$\alpha$.  (b) Contributions of the top and bottom quarks to the
$W$ and $Z$ self-energies.}
\label{fig1a}
\end{figure}
There is little shift in the $W$ mass, because that effect is already
absorbed into the observed value of the Fermi constant, so $\Delta
\hat{r}_W$ has no large $m_t$ dependence.  However, the $Z$ mass prediction
is shifted down.  In particular, the quantity $\hat{\rho}$ in
equation~(\ref{eq3}) depends quadratically on $m_t$.  It is given by
\cite{a9}
\beq \hat{\rho} \sim 1 + \rho_t, \label{rhohat} \eeq
where
\beq \rho_t = \frac{3G_F m_t^2}{8 \sqrt{2} \pi^2}  \sim
0.0031 (m_t/100 \ {\rm GeV})^2. \eeq
Here and throughout, $m_t$ refers to the pole mass, which corresponds
approximately to the kinematic mass relevant to direct searches
at colliders. (The running mass is discussed by Marciano in this
volume.) For $m_t$ in the range 100 -- 200~GeV the effect on
$\hat{\rho}$ can be quite significant.
$\rho_t$ propagates to other observables and generates most of the
$m_t$ dependence.  (The one important exception is the vertex
correction to $Z \ra b{\bar{b}}$ decay.)
  $\hat{\rho}$ contains additional contributions from
bosonic loops, including $\ln (M_H/M_Z)$ terms that
are strongly correlated with $\rho_t$.

{}From the precise value $M_Z = 91.1888 \pm 0.0044$~GeV from LEP one has
(using the expressions in~\cite{a13})
\beq \sin^2 \hat{\theta}_W (M_Z)= 0.2319 \pm 0.0005. \eeq
The uncertainty is an order of magnitude smaller than one had prior to the
$Z$-pole experiments at LEP.  The uncertainty from the experimental error
in the $Z$ mass is negligible, of order $0.00003$.  The theoretical
uncertainty $0.0003$ coming from $\Delta \hat{r}_W$ is much larger.  The
largest uncertainty, however, is from $m_t$ and $M_H$, $\sim 0.0004$.  Here
I have used the range of $m_t$ from the global best fit, and
60 GeV $< M_H <$ 1000 GeV.  If one knew $m_t$
one would have a more precise value of the weak angle.  The sensitivity is
displayed in Figure~\ref{fig1b}.  Clearly, one cannot determine the weak angle
from $M_Z$ alone because of the $m_t$ dependence.  One must have either
other indirect observables with a different dependence on $m_t$ or a direct
measurement.  Before discussing other possibilities, I will digress
somewhat on the radiative corrections~\cite{a8}.

\begin{figure}
\postbb{30 310 525 675}{/home/pgl/fort/nc/graph/mt/xxmt.ps}{0.8}
\caption[]{Values of $\sin^2 \hat{\theta}_W (M_Z)$ as a function of
$m_t$ from various observables.}
\label{fig1b}
\end{figure}

The radiative corrections fall into three categories.  First, there
are the reduced QED corrections, which involve the emission of real photons
and the exchange of virtual photons but do not include vacuum polarization
diagrams.  These constitute a gauge invariant set but depend on the
details of the experimental acceptances and cuts.  They generally are
removed from the data by the experimenters.  The second class has already
been described.  It is the electromagnetic vacuum polarization diagrams,
which lead to the running from $\alpha^{-1} \sim 137$ at low energies to
$\alpha (M_Z)^{-1} \sim 128$ at the $Z$-pole.  As we have seen this leads
to a significant uncertainty $\Delta \Delta \hat{r}_W \sim \Delta \alpha
(M_Z)/\alpha \sim 0.0009$~\cite{a8a},
which can lead to a shift of approximately 3~GeV
in the predicted value of $m_t$.

The electroweak corrections are now quite important.  One must include full
1-loop corrections as well as  dominant 2-loop effects.  The electroweak
corrections include and are dominated by the gauge self-energy diagrams for
the $W$, $Z$, and $\gamma Z$ mixing.  In addition, there are box diagrams
and vertex corrections, which are smaller but which have to be included.
Recently there has been some progress on the dominant 2-loop effects.  In
particular, the dominant terms of order $\alpha^2 m_t^4$ are included.  The
net effect is to replace (\ref{rhohat}) by~\cite{a11}
\beq \hat{\rho} \ra 1 + \rho_t \left[ 1 + \rho_t R \left(
\frac{M_H}{m_t} \right) \right], \eeq
where
$R$, which comes from 2-loop diagrams, is strongly dependent on
$M_H$, with $R (0) = 19 - 2 \pi^2$.  There are additional
smaller contributions which must be included in the numerical analysis.

There are also significant mixed QCD-electroweak diagrams, such as
those obtained
by the exchange of the gluon across the quarks in a self-energy diagram.
The dominant contribution
involves top quark loops and is of order $\alpha \alpha_s m_t^2$.
This leads to the replacement~\cite{a11a}.
\beq \hat{\rho} \ra 1 + \rho_t \left[ 1 - 2 \alpha_s (m_t)
\frac{\pi^2 + 3}{9 \pi} \right] \sim 1 + 0.9 \rho_t, \eeq
which raises the predicted value of $m_t$ by approximately 5\%.  Recently
there have been discussions and estimates of $t \bar{t}$ threshold
corrections, which are $O(\alpha \alpha_s^2 m_t^2)$.  These have been
estimated using both perturbative \cite{a12} methods and by dispersion
relations~\cite{a13}.  One estimate~\cite{a12} is that the effect is mainly
to shift the scale at which $\alpha_s$ should be evaluated for the $t$
quark loop, namely $\alpha_s (m_t) \ra \alpha_s (0.15 m_t)$.  This is in
good numerical agreement with the dispersion relation estimate,
which is used here. The threshold corrections raise
the predicted values of $m_t$ by $+ 3$~GeV.

\vglue 0.6cm
\subsubsection{\elevenit Renormalization of $\sin^2\theta_W$}
\vglue 0.4cm

There are a number of definitions of the renormalized weak angle
used in the literature, each with
its advantages and disadvantages.  At tree-level there are
several equivalent expressions, namely
\beq  \sin^2 \theta_W = \frac{g^{\prime2}}{g^2 + g^{\prime2}} = 1
- \frac{M_W^2}{M_Z^2} = \frac{\pi \alpha}{\sqrt{2} G_F M_W^2}.
\label{eq10a1} \eeq
The first definition is based on the coupling constants; the last two take
meaning only after spontaneous symmetry breaking has occurred, and
therefore mix in parts of the theory other than the gauge vertices.
At higher order one must define a renormalized angle.  One can use the
various expressions in equation~(\ref{eq10a1}) as starting points, and
the resulting definitions differ by finite terms of order $\alpha$, which
also depend on $m_t$ and $M_H$.  This has led to
considerable confusion (and heat).

Two common definitions are based on the spontaneous symmetry breaking (SSB)
of the theory, namely on the gauge boson masses.  The most famous is the
on-shell definition~\cite{onshell,a8}
\beq s_W^2 = 1 - \frac{M_W^2}{M_Z^2} = 0.2243 \pm 0.0012. \label{7.17}
\eeq
This is very simple conceptually.  However, the $W$ mass is not determined
as precisely as $M_Z$, so $s_W^2$ must actually be extracted from other
data and not from the defining relation (\ref{7.17}).  This leads to a strong
dependence on $m_t$, which accounts for almost all of the uncertainty in
$s_W^2$.  (The value given
for $s_W^2$ and the other definitions is from a global
fit to all data.)

In the on-shell scheme one defines the radiative correction parameter
$\Delta r$ by
\beq M^2_W = M^2_Z c^2_W = \frac{A^2}{s^2_W (1 - \Delta r)},  \label{delr}
   \eeq
where $ c^2_W = 1 - s^2_W $, $A$ is defined in (\ref{adef}), and
\beq \Delta r \sim 1 - \frac{\alpha}{\alpha(M_Z)} - \rho_t/\tan^2 \theta_W
  + {\rm small \ terms}, \eeq
which depends sensitively on $m_t$. For the allowed range, one expects
\beq \Delta r = 0.040 \pm 0.004 \pm 0.0009, \label{delrval} \eeq
where the second uncertainty is from $\alpha(M_Z)$.

The $Z$-mass definition \cite{a15},
\beq s^2_{M_Z} \left(1 - s_{M_Z}^2 \right) = \frac{\pi \bar{\alpha}
(M_Z)}{\sqrt{2} G_F M_Z^2} = 0.2312 \pm 0.0003, \eeq
is obtained by simply removing the $m_t$ dependence from the expression for
the $Z$ mass. $\bar{\alpha}^{-1}(M_Z) = 128.87 \pm 0.12$
differs by finite constants from the \msb quantity $\alpha^{-1}(M_Z)$.
This is the most precise -- the uncertainty is mainly from
$\alpha (M_Z)$.  The use of $s^2_{M_Z}$ is essentially equivalent to using
the $Z$ mass as a renormalized parameter, introducing the weak angle as a
useful derived quantity.  This scheme is simple and precise, and by
definition there is no $m_t$ dependence in the relation betwen $M_Z$ and
$s^2_{M_Z}$.  However, the $m_t$ dependence and uncertainties enter as soon as
one predicts other quantities in terms of it.

Both of the definitions based on spontaneous symmetry breaking are
awkward in the presence of any type of new physics that shifts the values of
the
gauge boson masses.  There are other definitions based on the gauge
coupling constants.  These are especially useful for applications to grand
unification, and they tend to be less sensitive to the presence of new
physics.  One is the modified minimal subtraction or $(\overline{MS})$
definition \cite{msbar,a8}
\beq \hat{s}^2_Z = \frac{\hat{g}'^2 (M_Z)}{\hat{g}'^2 (M_Z) +
\hat{g}^2 (M_Z)} = 0.2317 \pm 0.0004 ,\eeq
defined by removing the poles and associated constants from the gauge
couplings.  As we have seen, the uncertainty is mainly from $\alpha(M_Z)$
and $m_t$. It is useful to also define
$\hat{c}^2_Z = 1 - \hat{s}^2_Z$.
There are variant definitions\footnote{An
alternate form, $\hat{s}^2_{\rm ND}$~\cite{a17a}, which was used
frequently in earlier literature, does not decouple the $\ln m_t$ terms.
Its numerical value is close to the effective angle $\bar{s}^2_\ell$
for the favored $m_t$ range. The precise translations are given
in~\cite{a13,a7a}. Another variant, used in the program ZFITTER~\cite{a17aa},
is described
by Hollik in this volume.}
of $\hat{s}^2_Z$, depending on
the treatment of $\alpha \ln (m_t/M_Z)$ terms.  One cannot decouple all
such terms because $m_t \gg m_b$ breaks $SU_2$.  The version used here
\cite{a17,a13} decouples them from $\gamma - Z$ mixing, essentially eliminating
any $m_t$ dependence from the $Z$-pole asymmetry formulas.
The on-shell and \msb definitions are related by
\beq \hat{s}^2_Z = \kappa_W s^2_W,  \label{angles} \eeq
where $\kappa_W$ depends on $m_t$ and $M_H$. For example,
$\kappa_W$ = 1.033 for $(m_t,M_H)$ = (175,300) GeV, and the dominant $m_t$
dependence given by
\beq \kappa_W \sim 1 + \rho_t/\tan^2 \theta_W.  \label{onmsb} \eeq
The detailed relation is given in~\cite{a13,a17a}.

Finally, the experimental groups at LEP and SLC have made
extensive use of
\begin{eqnarray} \bar{g}_{Af} &=& \sqrt{\rho_f} t_{3f} \nonumber \\
\bar{g}_{Vf} &=& \sqrt{\rho_f} \left[ t_{3f} - 2 \bar{s}^2_f q_f \right].
\label{eq10a5} \end{eqnarray}
These are the effective axial and vector couplings of the $Z$ to fermion
$f$.  In equation (\ref{eq10a5}) $t_{3f} = \pm \frac{1}{2}$ is the weak
isospin of the left-handed component of
fermion $f$ and $q_f$ is its electric charge.  The electroweak
self-energy and vertex corrections are absorbed into the coefficient
$\rho_f$ and the effective weak angle $\bar{s}^2_f$. The $\bar{g}_{V,Af}$ are
obtained from the data after removing all photonic contributions.  In
principle there are also electroweak box contributions.  However, these are
very small at the $Z$ pole,
and are typically ignored or removed from the data.

There is a different effective weak angle for each type of fermion.
$\bar{s}^2_{f}$ is
related to the $\overline{MS}$ angle by
\beq \bar{s}^2_f = \kappa_f \hat{s}^2_Z, \eeq
where $\kappa_f$ is a form factor.  The best measured is for the charged
leptons. For the relevant $m_t$, $\kappa_\ell \sim 1.0013$~\cite{a17b},
so that
\beq \bar{s}^2_{\ell} \sim \hat{s}_Z^2 + 0.00028 = 0.2320 \pm
0.0004. \eeq
There is an additional theoretical uncertainty of $\pm 0.0001$ from
the precise definition of the angles and higher order effects.  These
effective angles are very simple for the discussion of the $Z$-pole data,
but are difficult to relate to other types of observables.  All
of these definitions have advantages and disadvantages, some of which are
listed in Table~\ref{tab3}.
Other definitions and schemes, such as the *-scheme, are described
by Hollik in this volume.

\begin{table} \centering  \small
 \begin{tabular}{|l|}  \hline \hline
On-shell : $s^2_W = 1 - \frac{M_W^2}{M_Z^2} = 0.2243\, (12)$ \\
\hline
$+$ most familiar \\
$+$ simple conceptually \\
$-$ large $m_t$ dependence from $Z$-pole observables \\
$-$ depends on SSB mechanism
$-$ awkward for new physics \\ \hline \hline
$Z$-mass : $s^2_{M_Z} = 0.2312 (3)$ \\ \hline
$+$ most precise (no $m_t$ dependence) \\
$+$ simple conceptually \\
$-$ $m_t$ reenters when predicting other observables \\
$-$ depends on SSB mechanism
$-$ awkward for new physics \\ \hline\hline
$\overline{MS}$ : $\hat{s}^2_Z = 0.2317\, (4)$ \\ \hline
$+$ based on coupling constants \\
$+$ convenient for GUTs \\
$+$ usually insensitive to new physics \\
$+$ $Z$ asymmetries $\sim$ independent of $m_t$ \\
$-$ theorists definition; not simple conceptually \\
$-$ usually determined by global fit \\
$-$ some sensitivity to $m_t$ \\
$-$ variant forms ($m_t$ cannot be decoupled in all processes; \\
\ \ \ $\hat{s}^2_{ND}$ larger by $0.0001 - 0.0002$) \\ \hline \hline
effective : $\bar{s}^2_{\ell} = 0.2320 \, (4)$ \\ \hline
$+$ simple \\
$+$ $Z$ asymmetry independent of $m_t$ \\
$+$ $Z$ widths: $m_t$ in $\rho_f$ only \\
$-$ phenomenological; exact definition in computer code \\
$-$ different for each $f$ \\
$-$ hard to relate to non $Z$-pole observables \\ \hline \hline
\end{tabular}
\caption[]{Advantages and disadvantages of several definitions of
the weak angle.}
\label{tab3}
\end{table}

\vglue 0.6cm
\subsubsection{\elevenit Other $Z$-Pole Observables}
\vglue 0.4cm

The other $Z$-pole observables can also be computed.  For example, the
partial width for $Z$ to decay into fermions $f \bar{f}$ is given
approximately by~\cite{a17c}
\beq \Gamma (f\bar{f}) \simeq C_f \frac{G_F M_Z^3}{6 \sqrt{2}
\pi} \left[ \left| \bar{g}_{Af} \right|^2 + \left| \bar{g}_{Vf}
\right|^2 \right].\label{eq12a1}  \eeq
For the heavier quarks and leptons kinematic mass corrections must be
applied, although they are only important for $\Gamma(\bar{b}b)$.
Effective couplings are proportional to $\sqrt{\hat{\rho}}$ so
that each partial width increases quadratically with $m_t$.  This comes
from the replacement
\beq \frac{M_Z g^2}{8 \cos^2 \hat{\theta}_W}  \ra \hat{\rho}
\frac{G_F}{\sqrt{2}} M_Z^3, \eeq
which incorporates many of the low energy corrections.  In
equation~(\ref{eq12a1}) there is an additional coefficient
\beq  C_f = \left\{ \begin{array}{cc} 1 + \frac{3 \alpha}{4 \pi}
q_f^2 \;\;\;\;  & ({\rm leptons}) \\ 3 \left( 1 + \frac{3
\alpha}{4 \pi} q_f^2 \right) & \left( 1 + \frac{\alpha_s}{\pi} +
1.409 \left( \frac{\alpha_s}{\pi} \right)^2  -12.77 \left(
\frac{\alpha_s}{\pi} \right)^3 \right) \ \ \ ({\rm quarks}),
 \end{array} \right.  \label{cfdef}
\eeq
which includes QED and QCD corrections. ($\alpha_s$ is
the strong coupling in the \msb scheme,
evaluated at $M_Z$.) For $\Gamma(\bar{b}b)$
there is additional $m_b$ and $m_t$ dependence in
the QCD corrections~\cite{a17d}.  The
$\alpha_s$ dependence of the hadronic widths leads to a determination of
$\alpha_s = 0.127 \pm 0.005$. (There
is also some sensitivity to the mixed QCD-electroweak contributions
to $\hat{\rho}$, which relates $M_Z$ to the other observables.) For fixed $M_Z$
most of the $m_t$ dependence is in the $\hat{\rho}$ factor.  One major
exception\footnote{There is also an indirect $m_t$ dependence in
$\bar{s}^2_f$ if one regards $M_Z$ as fixed.} is that $\Gamma (b \bar{b})$
decreases with $m_t$ due to special $m_t$-dependent vertex
corrections~\cite{a18}, \cite{a19}.  These are included in the $\rho_b$ and
$\kappa_b$ factors, but to an excellent numerical approximation
$\Gamma (b \bar{b})$ can be written
as \cite{a19},
\beq \Gamma (b\bar{b}) \ra \Gamma^0(b\bar{b}) \left( 1+ \delta^{SM}_{bb}
\right)\sim \Gamma^0 (b\bar{b}) \left[ 1 - 10^{-2}
\left( \frac{m_t^2}{2M_Z^2} - \frac{1}{5} \right) \right],\label{eq26a}
\eeq
where $\Gamma^0 (b\bar{b})$ is the standard model expression without the
corrections.  This special dependence is useful for separating
$m_t$ from the Higgs mass, and (especially) from such new physics
as higher-dimensional Higgs representations.

For describing the lineshape, it is convenient to use
the total width $\Gamma_Z$;
the hadronic peak cross section
$\sigma_{\rm had} = \frac{12 \pi}{M_Z^2} \; \frac{\Gamma(e
\bar{e}) \Gamma({\rm had})}{\Gamma_Z^2} ({\rm nb})$
(after removing QED effects); and the ratio
$R \equiv \Gamma({\rm had})
/\Gamma(\ell \bar{\ell})$ (where $ \Gamma(\ell \bar{\ell})$
is the average of the
$e, \mu,$ and $\tau$ widths after verifying lepton-family universality
and removing small $m_\tau$ effects). These are weakly
correlated, although in practice one generally includes the full
$M_Z$, $\Gamma_Z$, $\sigma_{\rm had}$, $R$, $A^{0\ell}_{FB}$
correlation matrix in fits\footnote{$M_W$ is also correlated to
$M_Z$ due to the theoretical uncertainty in $\alpha(M_Z)$.}.
$\Gamma_Z$ is sensitive to both $m_t$ and $\alpha_s$. Both
$R$ and $\sigma_{\rm had}$ are insensitive to $m_t$ because the
$\hat{\rho}$ factor cancels, while both (especially $R$) are
sensitive to $\alpha_s$.

The standard model predictions for $\Gamma_Z$, $\sigma_{\rm had}$,
and $R$
as a function of $m_t$ are compared with the experimental
results in Figures~\ref{gam1}.  ($\hat{s}^2_Z$
in $\bar{g}_{Vf} $ is obtained from $M_Z$).  One sees that
the agreement is excellent for $m_t$ in the 100 -- 200 GeV range.
The results of fits to the $Z$ widths are listed in Table \ref{stab}.
The prediction for $R_b$ is shown in Figure ~\ref{fig1}.
As already discussed, it is higher that the standard model prediction
for all allowed $m_t$, but favors smaller $m_t$.

The invisible width,
\bqa \Gamma({\rm inv}) & = & \Gamma_Z - \Gamma({\rm had}) - \sum_i
\Gamma(\ell_i \bar{\ell}_i ) \nonumber \\
 & \equiv & N_\nu \Gamma(\nu \bar{\nu}) \eqa
in Figure~\ref{gam2} is
clearly in agreement with $N_\nu = 3$ but not $N_\nu = 4$. In fact, the
result~\cite{a1} $N_\nu = 2.988 \pm 0.023$ not only eliminates extra
fermion families with $m_\nu \ll M_Z/2$, but also supersymmetric
models with light
sneutrinos ($\Delta N_\nu = 0.5$) and models with triplet ($\Delta N_\nu =
2$) or doublet ($\Delta N_\nu = 0.5$) Majorons \cite{numass}.
$N_\nu$ does not include
sterile ($SU_2$-singlet) neutrinos. However, the complementary bound
$N_\nu' < 3.3$ (95\% CL) from nucleosynthesis \cite{v27} {\it does}
include sterile neutrinos for a wide range of masses and mixings~\cite{numass},
provided their mass is less than $\sim 30$ MeV.

\begin{figure}
\small \def\baselinestretch{1} \normalsize
\postbb{70 45 570 710}{/home/pgl/fort/nc/graph/gam/xxgamf.ps}{0.8}
\caption[]{
The standard model predictions for $\Gamma_Z$, $\sigma_{\rm had}$,
and $R$ as a function of $m_t$, compared with the experimental
results.  For $\Gamma_Z$ the dotted, solid, and dashed lines are
for $M_H$ = 60, 300, and 1000 GeV, respectively.
The $M_H$ dependence is too small to see for the $\sigma_{\rm had}$
and $R$ graphs.  The QCD uncertainties are indicated.}
\label{gam1}
\end{figure}

\begin{figure}
\small \def\baselinestretch{1} \normalsize
\postbb{40 220 530 680}{/home/pgl/fort/nc/graph/gam/xxginv.ps}{0.6}
\caption[]{Theoretical prediction for $\Gamma({\rm inv}) $
in the standard
model with $N_\nu = 3$ and 4, compared with the experimental
value $499.8 \pm 3.5$ MeV.}
\label{gam2}
\end{figure}

In addition  there are various asymmetries observed at LEP and
SLD.  The forward-backward asymmetry for $e^+e^-
\ra Z \ra f \bar{f}$ is given, after removing photonic effects and
boxes, by
\beq A^{0f}_{FB} \simeq \frac{3}{4} A_e^0 A_f^0 , \eeq
where $A_f^0$ is defined in (\ref{eqn1}).
$A^{0\ell}_{FB}$ is the average of the $e$, $\mu$, and $\tau$ asymmetries
after verifying lepton-family universality. $A^{0b}_{FB}$, which is
determined after correcting for $B \bar{B}$ oscillations, is mainly
sensitive to $\hat{s}^2_Z$ (or small new physics effects) in $A^0_e$.
The LEP experiments also extract the weak angle from the
jet charge asymmetry~\cite{a1}.
Other asymmetries include the polarization of produced $\tau$'s.
The polarization is given as a function of the scattering
angle $z \equiv \cos \theta$ by
\beq
P_\tau^0 = - \frac{A^0_\tau + A^0_e \frac{2 z}{1+z^2}}{1
  + A^0_\tau A^0_e \frac{2 z}{1+z^2}}.  \eeq
{}From the angular distribution one can obtain
$A_{\tau}^0$ and $A_e^0$ with little correlation, with
$A_{\tau}^0$ coming mainly from the average polarization and $A_e^0$ mainly
from its forward-backward asymmetry.  The SLD collaboration has polarized
electrons; from the left-right asymmetry as the polarization is reversed
one can also determine $A_e^0$, namely $A_{LR}^0 = A_e^0$.

All of these asymmetries are independent of $m_t$ when expressed in terms
of the effective angles $\bar{s}_f^2$ and almost independent of $m_t$ when
expressed in terms of the $\overline{MS}$ angle $\hat{s}^2_Z$.  One can
therefore determine $\bar{s}_\ell^2$ or $\hat{s}^2_Z$ from the data without
theoretical uncertainties from $m_t$.  On the other hand, in the on-shell or
$Z$-mass schemes the formulas involve quadratic $m_t$ dependence.

The predictions for $A^{0\ell}_{FB}$,
$A^0_{\ell}$, and $A^{0b}_{FB} $ are compared with the experimental
data in Figure~\ref{asym}. Again, the agreement is excellent.

\begin{figure}
\small \def\baselinestretch{1} \normalsize
\postbb{60 40 570 720}{/home/pgl/fort/nc/graph/afb/xxafb.ps}{0.8}
\caption[]{Theoretical prediction for $A^{0\ell}_{FB}$,
$A^0_{\ell}$, and $A^{0b}_{FB} $ in the standard
model as a function of $m_t$ for $M_H$ = 60 (dotted line), 300 (solid), and
1000 (dashed) GeV, compared with the experimental values. The
theoretical uncertainties from $\Delta \Delta
\hat{r}_W = \pm 0.0009$ are also indicated. For $A^0_{\ell}$
the value $0.139 \pm 0.007$ is the average of the LEP values from
$A_e^0 (P_\tau)$ and $A_\tau^0 (P_\tau)$, while $0.1637 \pm 0.0075$ is the
SLD value from $A_{LR}^0$.}
\label{asym}
\end{figure}

The values obtained for \sinn in the \msb and on-shell schemes
from various $Z$-pole observables, $M_W$, and low
energy neutral current processes are listed in Table~\ref{stab}
and displayed in Figure~\ref{sq}.
The low energy values are not as precise as those from
the $Z$-pole and $M_W$. However, they are still important
as they probe different couplings and kinematic ranges, and
are sensitive to certain types of new physics to which the $W$
and $Z$ are blind.

\begin{table}\centering  \small \def\baselinestretch{1} \normalsize
\begin{tabular}{|c|c|c|} \hline
Data & $\hat{s}^2_Z$ & $s^2_W $  \\ \hline
$M_Z$ & 0.2319 \ppm 0.0005 & 0.2245 \ppm 0.0013 \\
$M_W, \frac{M_W}{M_Z}$ & 0.2326 \ppm 0.0011 & 0.2251 \ppm 0.0014 \\
$ \Gamma_{Z}, \sigma_{\rm had},R $ &
0.2317 \ppm 0.0018 & 0.2242 \ppm 0.0017 \\
$A^{0\ell}_{FB}$ & 0.2308 \ppm 0.0009 & 0.2234 \ppm 0.0013 \\
$A^0_{\tau} \left(P_\tau \right),
A^0_e \left(P_\tau \right)$ &  0.2322 \ppm 0.0009 & 0.2248 \ppm 0.0013 \\
$A^{0b}_{FB}$ & 0.2324 \ppm 0.0007 & 0.2249 \ppm 0.0012 \\
all LEP asymmetries & 0.2319 \ppm 0.0004 & 0.2242 \ppm 0.0011 \\
$A^0_{LR}$ & 0.2291 \ppm 0.0010 & 0.2218 \ppm 0.0013 \\
$\nu_\mu(\bar{\nu}_\mu) N \ra \nu_\mu(\bar{\nu}_\mu) X$
& 0.234 \ppm 0.005 & 0.226 \ppm 0.005 \\
$\nu_\mu(\bar{\nu}_\mu) p \ra \nu_\mu(\bar{\nu}_\mu) p$
& 0.212 \ppm 0.032 & 0.205 \ppm 0.030 \\
$\nu_\mu(\bar{\nu}_\mu) e \ra \nu_\mu(\bar{\nu}_\mu) e$
& 0.228 \ppm 0.008 & 0.221 \ppm 0.007 \\
atomic parity & 0.223 \ppm 0.008 & 0.216 \ppm 0.008 \\
$e^{\uparrow \!\downarrow} D \ra eX$ & 0.223 \ppm 0.018 & 0.216 \ppm 0.017 \\
All & 0.2317 \ppm 0.0004 & 0.2243 \ppm 0.0012 \\
\hline
\end{tabular}
\caption[]{Values of $\hat{s}^2_Z$ (\msb) and $s^2_W $ (on-shell)
obtained from various inputs, assuming the global best fit values
$m_t = 175 \pm 11$ GeV (for $M_H = 300$ GeV) and
$\alpha_s = 0.127 \pm 0.005$, correlated with
60~GeV $< M_H < 1000$~GeV.
For $\Gamma_Z$ and $\sigma_{\rm had}$
the experimental $M_Z$ is used, so that
\sinn is determined from the electroweak vertices.
(The uncertainty in $\hat{s}^2_Z$ would be reduced to 0.0004 if one used
the theoretical formulas, but the additional sensitivity
is mainly due to the $M_Z^3$ factor in $\Gamma_Z$.)
For deep inelastic scattering
($\nu_\mu(\bar{\nu}_\mu)N \ra \nu_\mu(\bar{\nu}_\mu)X$)  from
(approximately) isoscalar targets (Perrier, this volume)
the uncertainty includes 0.003 (experiment) and 0.005 (theory).
$\nu_\mu(\bar{\nu}_\mu)p$ and $\nu_\mu(\bar{\nu}_\mu)e$ refer
respectively to elastic scattering from nucleons (Mann, this volume)
and electrons (Panman, this volume).
$e^{\uparrow \!\downarrow} D$ refers to the SLAC polarized $eD$
asymmetry (Souder, this volume).
For atomic parity violation (Masterson and Wieman, this volume),
the experimental and
theoretical (Blundell, Johnson, and Sapirstein, this volume)
components of the error are 0.007 and 0.004 respectively.}
\label{stab}
\end{table}

\begin{figure}
\small \def\baselinestretch{1} \normalsize
\postbb{60 250 520 670}{/home/pgl/fort/nc/graph/xq/xxxq.ps}{0.6}
\caption[]{\shz \ obtained from various observables
assuming $m_t = 175 \pm 11 $~GeV, $\alpha_s = 0.127 \pm 0.005$, and
60~$< M_H < 1000$~GeV.}
\label{sq}
\end{figure}

\vglue 0.6cm
\subsection{\elevenit The Standard Model Parameters:
$m_t$, $\alpha_s$, $\sin^2\theta_W$}
\vglue 0.4cm

($m_t$, and $\sin^2\theta_W$):
There are now sufficiently many observables that one can
precisely determine $\hat{s}^2_Z$, $m_t$, and $\alpha_s (M_Z)$
simultaneously.  For example, $\hat{s}^2_Z$ can be determined
from the asymmetries,  $m_t$ from the $W$ and $Z$ masses, and
$\alpha_s (M_Z)$ from the hadronic $Z$-widths.  In practice all
of these quantities are determined from a simultaneous fit.  The
results of fits to various sets of data are shown in
Table~\ref{tab4}.
\begin{table} \centering
\begin{tabular}{|ccccc|}  \hline \hline
Set & $\hat{s}^2_Z$ & $\alpha_s (M_Z)$ & $m_t$ (GeV) & $\Delta
\chi^2_H$ \\ \hline
All indirect & $0.2317 (3)(2)$ & $0.127 (5)(2)$ & $175 \pm
11^{+17}_{-19}$ & 4.4 \\
Indirect $+$ CDF (174 $\pm$ 16)    & $0.2317 (3)(3)$ & $0.127
(5)(2)$ & $175\pm 9^{+12}_{-13}     $ & 4.4 \\
Indirect $+\alpha_s^{\rm other}
(0.116 \pm 0.005)$    & $0.2316 (3)(2)$ & $0.122
(3)(1)$ & $178^{+10 \; +17}_{-11 \; -19}    $ & 6.0 \\
LEP $+$ low energy  & $0.2320 (3)(2)$ & $0.128 (5)(2)$ &
$168^{+11 \; +17}_{-12 \; -19}$ & 2.7 \\
All indirect $(S=2.2)$ & $0.2319 (3)(2)$ & $0.128 (5)(2)$ &
$170^{+11 \; +17}_{-12 \; -19}$ & 3.3 \\
$Z$-pole               & $0.2316 (3)(1)$ & $0.126 (5)(2)$ &
$179^{+11 \; +17}_{-12 \; -19}$ & 4.2 \\
LEP                    & $0.2320 (4)(2)$ & $0.128 (5)(2)$ &
$170^{+12 \; +18}_{-13 \; -20}$ & 2.6 \\
SLD $+ \; M_Z$      & $0.2291 (10)(0)$ & ---            &
$251^{+24 \; +21}_{-26 \; -23}$ & \   \\ \hline
\end{tabular}
\caption[]{Results for the electroweak parameters in the standard model from
various sets of data.  The central values assume $M_H = 300$~GeV, while the
second errors are for $M_H \ra 1000 \ (+)$ and $60 \ (-)$.  The last column is
the increase in the overall $\chi^2$ of the fit as $M_H$ increases from 60
to 1000. From \cite{a7c}.}
\label{tab4}
\end{table}
The first row of the table includes the global fit to all indirect
data\footnote{The correlation coefficients are
$\rho_{\hat{s}^2_Z m_t} = -0.67$,
$\rho_{\hat{s}^2_Z \alpha_s} = 0.30$, and $ \rho_{\alpha_s m_t}
=-0.20$.
The overall $\chi^2$ of the fit
is 181 for 206 d.f., which is low (mainly due to the older
neutral current data) but acceptable: the probability of
$\chi^2 \le 181$ is $10\%$. The correlations for the other data
sets are similar.}.
The predicted value,
\beq  m_t = 175 \pm 11^{+17}_{-19} \ {\rm GeV}, \label{smmt} \eeq
is in remarkable agreement with the value $174
\pm 16$~GeV suggested by the CDF candidate events \cite{a2b}.
The second
row includes the direct (CDF) value for $m_t$ as a separate constraint.
Since the indirect data is consistent with the CDF value, adding
it has little effect on the standard model fits. (It is very important
for the beyond the standard model fits, however.) The third row
includes the value
$\alpha_s^{\rm other} = 0.116 \pm 0.005$ of $\alpha_s(M_Z)$
obtained from jet and low energy data \cite{a22} as a separate
constraint. The resulting $\alpha_s = 0.122(3)(1)$ may be viewed
as a weighted average of these other determinations with the lineshape
value of $0.127(5)(2)$.
The other fits show the sensitivity to the various data sets.  The fourth
row includes the LEP results and the low energy data but not SLD.
Comparing with the first row one sees that the predicted $m_t$ is pulled up
significantly (by $\sim 7$~GeV) by the SLD result.  The next row combines
the LEP and SLD measurements of $A_e$, increasing the error in the
weighted average by the scale factor\footnote{$S$ is the square root
of the $\chi^2/df$. This is the procedure recommended by the Particle
Data Group~\cite{PDGSF} when there is a discrepancy between
experiments.} $S = 2.2$.  The
last rows are the result of the $Z$-pole, LEP, and SLD observables by
themselves.

The central value in (\ref{smmt}) is for $M_H = 300$ GeV, while
the second uncertainty is from $M_H$ varying from 60 to 1000 GeV.
This reflects the strong correlation between the $m_t^2$ and
$\ln(M_H)$ terms in the $\hat{\rho}$ parameter (but not in
the $Zb\bar{b}$ vertex corrections). For other values of $M_H$ one
finds the approximate prediction
\beq m_t \sim 175 \pm 11 + 13 \ln \left( \frac{M_H}{300 \rm
GeV} \right), \label{mtmh} \eeq
where the coefficient 13 is obtained from the fits (see Figure
\ref{ht}) and should
be viewed as an approximate interpolation of
more complicated formulae. In particular, supersymmetric extensions
of the standard model, which involve a light standard model-like Higgs,
favor the lower part of the range in (\ref{smmt}).

\begin{figure}
\small \def\baselinestretch{1} \normalsize
\postbb{70 220 540 640}{/home/pgl/fort/nc/graph/mtmh/xxmtmh.ps}{0.55}
\caption[]{Best fit value for $m_t$ and upper and lower limits as a
function of $M_H$. The direct lower limit $M_H >$ 60 GeV \cite{v8} and
the approximate triviality limit \cite{a25} $M_H < $ 600 GeV are also
indicated. The latter becomes $M_H < $ 200 GeV if one requires that the
standard model holds up to the Planck scale. The CDF value
$m_t = 174 \pm 16$ GeV and the D0 bound $m_t > 131$ GeV are also
indicated.}
\label{ht}
\end{figure}

The $\chi^2 $ distributions for the fit of the indirect data as
a function of $m_t$ are shown for various values of $M_H$ in
Figure~\ref{chis}. One again observes the strong correlation
between $m_t$ and $M_H$.
{}From the indirect data one can obtain upper and lower limits
on $m_t$. The weakest upper limit is for $M_H$ = 1000 GeV,
from which one finds $m_t < 205 \ (209)$ GeV at 90 (95)\% CL.
The corresponding limits for other $M_H$ are 170 (174) GeV
for $M_H$ = 60 GeV and 188 (192) GeV for $M_H$ = 300.
Similarly, the indirect data alone set significant lower
limits on $m_t$, which would continue to hold even in the presence
of nonstandard $t$ decays which could invalidate the direct collider
searches. The weakest limit is for $M_H$ = 60 GeV, for which one
obtains $m_t > 140 \ (135) $ GeV at 90 (95)\% CL.

\begin{figure}
\small \def\baselinestretch{1} \normalsize
\postbb{50 190 520 680}{/home/pgl/fort/nc/graph/chis/xxchis.ps}{0.50}
\caption[]{$\chi^2$ distribution
for all indirect data (206 df) in the standard model
 as a function of $m_t$, for $M_H = 60$, \mz, 300, and
1000~GeV. The direct constraints on $m_t$ from CDF and D0 are
displayed but are  not
included in the $\chi^2$.} \label{chis}
\end{figure}

{}From the indirect data one obtains
\beqa \hat{s}^2_Z & =
& 0.2317\pm 0.0003 \pm 0.0002 \nonumber \\
 s^2_W & = & 0.2243 \pm 0.0012 \nonumber \\
 \bar{s}^2_\ell & = & 0.2320 \pm 0.0003 \pm 0.0002
 \label{eq54} \eeqa
for the the \msb, on-shell, and effective weak angles, respectively.
The first uncertainties  are mainly from $m_t$ and $\alpha(M_Z)$,
while the second is from
$M_H$ in the range 60 - 1000~GeV.
$\hat{s}^2_Z$ and $\bar{s}^2_\ell$
are much less sensitive to $m_t$ and $M_H$ than
$s^2_W$.  With the exception of $A^0_{LR}$
the values obtained from individual observables
are in excellent agreement with (\ref{eq54}).  In particular, the
$\hat{s}^2_Z$ values obtained assuming $m_t =
175 \pm 11$~GeV, $\alpha_s = 0.127 \pm 0.005$,
and 60~GeV~$<M_H < 1000$~GeV are shown in
Table~\ref{stab} and in Figure~\ref{sq}.  The agreement is
remarkable.

One can also extract the radiative correction parameter
$\Delta \hat{r}_W$ defined in (\ref{eq2}). Fitting to all
indirect data and keeping the full $m_t$ dependence in $\hat{\rho}$
but leaving $\Delta \hat{r}_W$ free, one finds
\beq    \Delta \hat{r}_W = 0.067 \pm 0.002, \eeq
compared with the expectation $0.0706(2)(9)$ in (\ref{delrhat}).
Including the CDF $m_t$ value, $\Delta \hat{r}_W = 0.068 \pm 0.002$.
For the analogous parameter
$\Delta r$ in the on-shell scheme, eqn. (\ref{delr}), one obtains
\beq \Delta r = 0.044 \pm 0.005 \label{eq55} \eeq
from the indirect data, or $0.041 \pm 0.003$ including CDF,
compared to the expectation $0.040(4)(1)$ in (\ref{delrval}).
\vglue 0.4cm

\noindent ($\alpha_s$):
Using the results of the 1993 LEP energy scan one can now extract the strong
coupling constant $\alpha_s$ at the $Z$-pole with a small experimental
and theoretical error,
\beq \alpha_s (M_Z) = 0.127 \pm 0.005 \pm 0.002 \;\;\;\; {\rm
(lineshape)},\label{eq30a} \eeq
where the second uncertainty is from $M_H$. $\alpha_s$ is only weakly
correlated with the other parameters.  It is determined mainly from the
ratio $R \equiv \Gamma (\rm had)/ \Gamma (\ell \bar{\ell})$, which is
insensitive to $m_t$ (except in the $b \bar{b}$ vertex). There is
also sensitivity from $\Gamma_Z$, $\sigma_{\rm had}$, and in the
mixed QCD-electroweak corrections which relate $M_Z,\ \hat{s}^2_Z$, and $M_W$.
This determination is very clean theoretically, at least
within the standard model.  It is the $Z$-pole version of the long held
view that the ratio of hadronic to leptonic rates in $e^+e^-$ would be a
``gold-plated'' extraction of $\alpha_s$ and test of QCD.  Using a recent
estimate \cite{a21} of the $(\alpha_s/\pi)^4$
corrections to $C_f$ in (\ref{cfdef}), {\it
i.e.} $- 90 (\alpha_s/\pi)^4$, one can estimate that higher-order terms
lead to an additional uncertainty $\sim \pm 0.001$ in the $\alpha_s(M_Z)$
value in (\ref{eq30a}).  It should be cautioned, however, that the
lineshape value is sensitive to the presence of new
physics which affects the hadronic $Z$ decays. In particular, it will be
discussed in Section \ref{zbbsec} that if one allows for new
physics in the $Zb\bar{b}$ vertex to account for $R_b$, the extracted
value of $\alpha_s$ decreases to $0.111(9)(1)$.

The lineshape value of $\alpha_s$ is an excellent agreement with the
independent value $\alpha_s (M_Z) = 0.123 \pm 0.006$ extracted from jet
event shapes at LEP using resummed QCD \cite{a22}.  It is also in
excellent agreement with the prediction \cite{a22aa}
\beq \alpha_s (M_Z) \sim 0.129 \pm 0.008 ,\;\;\;\;\; {\rm
SUSY-GUT} \eeq
of supersymmetric grand unification.  As can be seen in Table~\ref{tab5},
however, it is somewhat larger than some of the low energy determinations
of $\alpha_s$ (which are then extrapolated theoretically to the $Z$-pole),
in particular those from deep inelastic scattering, the lattice
calculation of the charmonium spectrum\footnote{The value
$0.110 \pm 0.006$ \cite{a22a} has increased somewhat from the published value
of $0.105 \pm 0.004$ \cite{a22b}, reducing the discrepancy.},
and a recent lattice calculation of the bottomonium spectrum which claims
a very small uncertainty \cite{a22c}.
This has led
some authors to speculate that there might be a light gluino which would
modify the running of $\alpha_s$~\cite{gluino}.
It should be noted, however, that there is an
independent low energy LEP determination from the ratio $R_\tau$ of
hadronic to leptonic $\tau$ decays, which gives a larger value.

The third row of Table \ref{tab4} includes the value $\alpha_s^{\rm other}
= 0.116 \pm 0.005$ obtained from low energy and jet data \cite{a22}.
The resulting value $\alpha_s = 0.122(3)(1)$ can be regarded as
a weighted fit to all data, including the $Z$ lineshape. However, given
the discrepancies between the individual determinations, caution is
advised in accepting either this value or the small error. For this reason,
when discussing grand unification in Section \ref{gutsec}, $\alpha_s$
will generally be taken as a prediction rather than an input, or else
the more conservative range $0.12 \pm 0.01$ will be used.

\begin{table}                       \centering
\begin{tabular}{|l|c|} \hline \hline
Source & $\alpha_s (M_Z)$ \\ \hline
$R_\tau = \Gamma(\tau \RA \nu_\tau + {\rm had})/\Gamma(\tau \RA
  {\rm leptons})$ & $0.122 \pm 0.005$ \\
Deep inelastic & $0.112 \pm 0.005$ \\
$\Upsilon$, $J/\Psi$ decays & $ 0.113 \pm 0.006$ \\
Charmonium spectrum (lattice) & $0.110 \pm 0.006$ \\
Bottomonium spectrum (lattice) & $0.115 \pm 0.002$ \\
LEP, event topologies & $0.123 \pm 0.006$ \\
LEP, lineshape & $0.127 \pm 0.005 \pm 0.002$ \\
\hline
\end{tabular}
\caption[]{Values of $\alpha_s$ at the $Z$-pole extracted from
various methods.}
\label{tab5}
\end{table}

The value of \alsz \ from the precision experiments is
anticorrelated with \mt, as can be seen in Figure \ref{alsmt}.
In particular, larger \mt \ corresponds to slightly smaller \alsz, in
better agreement with the low energy data. As mentioned above,
one would also obtain a smaller value if there is new physics
which enhances the $Zb\bar{b}$ vertex.
\begin{figure}
\small \def\baselinestretch{1} \normalsize
\postbb{70 220 540 640}{/home/pgl/fort/nc/graph/alsmt/xxalsmt.ps}{0.55}
\caption[]{90\% CL allowed region in \alsz \ and \mt \ from a combined
fit to precision $Z$-pole and other data (but not including event
topology and low energy determinations of \alsz).}
\label{alsmt}
\end{figure}

\vglue 0.6cm
\subsection{\elevenit The Higgs Mass}
\vglue 0.4cm

One can attempt to use the precision
data to constrain  the Higgs boson mass.  This enters
$\hat{\rho}$ logarithmically and is strongly correlated with the
quadratic $m_t$   dependence in everything but  the $Z \ra b \bar{b}$
vertex correction.
The $\chi^2$ distributions as a function of the Higgs
mass are shown  with and without the additional CDF constraint
$m_t = 174 \pm 16$ GeV in Figure~\ref{fig2}.
In both cases, the minimum occurs at or near the lower limit,
60 GeV, allowed by direct searches at LEP. (The increase in $\chi^2$
for $M_H = 1000 $ GeV is shown in Table~\ref{tab4}.)
A low value for $M_H$ is consistent with the minimal
supersymmetric extension of the standard model, which generally predicts a
relatively light standard model-like Higgs scalar.  However, the constraint
is weak statistically.  From the $\chi^2$ distribution one obtains the
weak upper limits
\beq {\rm indirect: \;\;\;} M_H < 570 \ (880) \ {\rm GeV} \eeq
at 90 (95)\% CL from the indirect precision data, and
\beq {\rm indirect+CDF\; : \;\;\;} M_H < 510 \ (730) \ {\rm GeV} \eeq
including the CDF direct constraint from $m_t$.  (These results
include the direct LEP limit $M_H > 60$~GeV~\cite{a25}.)
Clearly, no definitive conclusion can be drawn.
\begin{figure}
\postbb{15 40 560 710}{/home/pgl/fort/nc/graph/chis/xxhiggs2g.ps}{0.8} 
\caption[]{ Increase in $\chi^2$ from the best fits as a
function of $M_H$, with and without the CDF constraint $m_t = 174 \pm 16$ GeV,
and distributions omitting $R_b$ and/or $A^0_{LR}$.}
\label{fig2}
\end{figure}
Furthermore, the sensitivity to $M_H$ is driven almost entirely by
the experimental values of $R_b$ and $A^0_{LR}$, both of which are
well above the standard model expectations. Omitting these leads
to an almost flat $\chi^2$ distribution, as can be seen in Figure~\ref{fig2}.
If these are due to
large statistical fluctuations or new physics the constraint on
$M_H$ essentially disappears.

The weak $M_H$ dependence does not imply that the data is insensitive to
the spontaneous symmetry breaking mechanisms.  Alternative schemes
generally yield large effects on the precision observables,
as will be described below.

\vglue 0.6cm
\subsection{\elevenit Have Electroweak Corrections Been Seen?}
\vglue 0.4cm

The data can also be interpreted in terms of whether one has actually
observed the electroweak (as opposed to the simple running $\alpha$)
corrections.  Novikov {\it et al}. \cite{a15} have noted that there is a
large cancellation between the fermionic and bosonic contributions to the
$W$ and $Z$ self-energies, and that until recently the data
could actually be fit by a properly interpreted Born theory.  However, the
data is now sufficiently good that even given the cancellations these
electroweak loops are needed at the $2\sigma$ level.  Gambino and Sirlin
\cite{a23} and Dittmaier {\it et al.} \cite{a24} have interpreted the data in
somewhat different way.  They have argued that the fermionic loops,  both
in the running of $\alpha$ and the $t,b$ loops, are unambiguous
theoretically, and certainly should be there if the theory is to make any
sense. However, the bosonic loops, which involve triple-gauge vertices,
gauge-Higgs vertices, etc., have never been independently tested in other
processes.  They have shown that the data are inconsistent if one simply
ignores bosonic loops (which are a gauge-invariant subset of diagrams),
thus providing convincing though indirect evidence
for their existence.

\vglue 0.6cm
\section{\elevenbf Model Independent Analyses}
\vglue 0.4cm

Long before the LEP era, the standard model was strongly favored
compared to competing gauge theories by model independent analyses of
neutrino scattering, atomic parity violation, and $e^+e^-$ below the
$Z$ pole \cite{nuq}.
In a model independent analysis one writes an
effective \lag\ involving all of the four-fermi operators that can be
obtained at tree-level in a gauge theory\footnote{In practice one
usually assumes family universality.}, and then tries to determine their
coefficients directly from the data.  Each electroweak gauge theory
makes a prediction for the coefficients.  One can therefore see
whether the coefficients are uniquely determined and whether they are
in agreement with the standard model predictions.

The model independent analyses were important historically in
establishing the uniqueness of the standard model. They are
still important today because the four-fermi parameters that they
determine are not quite the same as those measured in the
(more accurate) $Z$-pole experiments, at least in the presence
of new physics. The former
are sensitive to all standard model and new physics
contributions to the process. The latter are the actual
couplings of the $Z$ to the corresponding fermions, and are
sensitive only to the types of new physics which directly affect
the $Z$.

\begin{table} \centering
\small \def\baselinestretch{1} \normalsize
\begin{tabular}{|ll|} \hline \hline
Quantity & Standard Model Expression \\ \hline
$\epsilon_L(u)$  &  $\rho^{NC}_{\nu N} \left( \frac{1}{2} -
\frac{2}{3} \kappa_{\nu N} \sin^2 \theta_W + \lambda_{uL}\right) $\\
$\epsilon_L(d)$  &  $\rho^{NC}_{\nu N} \left(-\frac{1}{2} +
\frac{1}{3} \kappa_{\nu N} \sin^2 \theta_W + \lambda_{dL} \right)$ \\
$\epsilon_R(u)$  &  $\rho^{NC}_{\nu N} \left(  -
\frac{2}{3} \kappa_{\nu N} \sin^2 \theta_W +\lambda_{uR} \right)$ \\
$\epsilon_R(d)$  &  $\rho^{NC}_{\nu N} \left(
\frac{1}{3} \kappa_{\nu N} \sin^2 \theta_W + \lambda_{dR} \right) $\\
$g^{\nu e}_V$  &  $\rho_{\nu e} \left( - \frac{1}{2} + 2 \kappa_{\nu e}
\sin^2 \theta_W \right)$ \\
$g^{\nu e}_A$ & $\rho_{\nu e} \left(-\frac{1}{2} \right)$ \\
$C_{1u}$ & $\rho'_{eq} \left(-\frac{1}{2} + \frac{4}{3} \kappa'_{eq}
\sin^2 \theta_W \right)$ \\
$C_{1d}$ & $\rho'_{eq} \left(\frac{1}{2} - \frac{2}{3} \kappa'_{eq}
\sin^2 \theta_W \right)$ \\
$C_{2u}$ & $\rho_{eq} \left(-\frac{1}{2} + 2 \kappa_{eq}
\sin^2 \theta_W \right) + \lambda_{2u} $ \\
$C_{2d}$ & $\rho_{eq} \left(\frac{1}{2} - 2 \kappa_{eq}
\sin^2 \theta_W \right) + \lambda_{2d} $ \\ \hline
\end{tabular}
\caption[]{Standard model expressions for the neutral-current parameters
for $\nu$-hadron, $\nu e$, and $e$-hadron processes.  If radiative
corrections are ignored, $\rho = \kappa = 1, \; \lambda = 0$.  At $O
(\alpha)$ in the on-shell scheme, $\rho^{NC}_{\nu N} = 1.0089$,
$\kappa_{\nu N} = 1.0356$, $\lambda_{uL} = - 0.0032$, $\lambda_{dL} =
-0.0026$, and $\lambda_{uR} = 1/2 \lambda_{dR} = 3.6 \x 10^{-5}$ for
$m_t = 175$~GeV, $M_H = 300$~GeV, $M_Z = 91.1888$ GeV,
and $\langle Q^2 \rangle = 20$~GeV$^2$.  For $\nu
e$ scattering,
$\rho_{\nu e} = 1.0137 $ and
$\kappa_{\nu e} = 1.0358$
(at $\langle Q^2 \rangle = 0)$.  For atomic parity violation,
$\rho'_{eq} = 0.9880$ and $\kappa'_{eq} = 1.034$.  For the SLAC
polarized electron experiment, $\rho'_{eq} = 0.979$, $\kappa'_{eq} =
1.032$, $\rho_{eq} = 1.001$, and $\kappa_{eq} = 1.06$ after
incorporating additional QED corrections, while $\lambda_{2u} =
-0.013$, $\lambda_{2d} = 0.003$. The dominant $m_t$ dependence
is given by $\rho \sim 1 + \Delta \rho_t$, while $\kappa
\sim 1 + \Delta \rho_t/\tan^2 \theta_W$ (on-shell) or
$\kappa \sim 1$ (\msb).}
\label{tab10}
\end{table}

Previous to LEP the most precise neutral current tests involved the
neutrino-quark interactions, as described in the article by
Perrier and in~\cite{nuq,nuq2}.  There have been deep
inelastic experiments on (approximately)
isoscalar and $p$ and $n$ targets, $\nu N \ra
\nu X, \; \nu p \ra \nu X, \; \nu n \ra \nu X$, elastic scattering such
as $\nu p \ra \nu p$, and various inelastic reactions, such as
coherent $\nu N \ra \nu \pi^0 N$.  In particular, the deep inelastic
cross sections for neutral current scattering divided by the
corresponding charged current cross sections for targets such as iron
and carbon, for which most of the strong interaction
uncertainties\footnote{The largest residual uncertainty is in the charm
quark threshold in the charged current
denominator~\cite{a6,nuq,mcth}.}~\cite{lls}
and those involving the neutrino flux cancel in the ratio, have been
measured at the 1\% level.  The less precise measurements on proton
and neutron targets constrain the isospin structure of the
current.  Assuming family universality and left-handed neutrinos, the
most general effective four-fermi interaction for $\nu q$ scattering
that can be generated from gauge interactions, \ie allowing only
vector and axial interactions, is\footnote{Alternate parametrizations
are described in~\cite{kim}.}
\beq -L^{\nu N} = \frac{G_F}{\sqrt{2}} \bar{\nu} \gamma^\mu (1 -
\gamma^5) \nu    \sum_{i = u,d} \left[ \epsilon_L (i) \bar{q}_i
\gamma_\mu (1 - \gamma^5) q_i + \epsilon_R (i) \bar{q}_i \gamma_\mu (1
+ \gamma^5) q_i \right], \label{eqch122} \eeq
where the parameters $\epsilon_L(i)$ and $\epsilon_R(i)$ refer to the
interactions of left- and right-handed quark $i$ with neutrinos.  The
standard model expressions for the $\epsilon$'s are listed, including
the radiative corrections, in Table~\ref{tab10}.
(Specific values of the radiative corrections are given in the on-shell
scheme, which was used in most of the analyses~\cite{nuq}. They
can be translated to the \msb scheme using (\ref{angles}).)
Other gauge theories would give other values.
The data is sufficient to determine the four
$\epsilon$'s uniquely\footnote{The deep inelastic experiments usually
presented their results in terms of \sinn only and required
considerable reanalysis for the model independent studies \cite{nuq}.}.
The results are shown in Figure~\ref{fig45}.
The left-hand couplings give the most precise
low energy determination of $\sin^2
\theta_W$.  The values of the parameters are listed
along with the standard model predictions in
Table~\protect\ref{tab11}.  Also given are the values and predictions
for the quantities $g^2_L$, $g^2_R$, $\theta_L$, $\theta_R$, which are
the squares of the radii and the angles in the $\epsilon_L$ and
$\epsilon_R$ planes.  These quantities are much more weakly correlated
than the $\epsilon$'s themselves.

\begin{figure}
\small \def\baselinestretch{1} \normalsize
\postbb{120 40 470 710}{/home/pgl/fort/nc/graph/modind/xxnq.ps}{0.6}
\caption[]{The allowed regions in $\epsilon_L(u)$, $\epsilon_L(d)$,
$\epsilon_R(u)$ and $\epsilon_R(d)$, compared with the standard model
predictions as a function of $s^2_W$.   The agreement is excellent.}
\label{fig45}
\end{figure}

\begin{table} \centering
\small \def\baselinestretch{1}
\begin{tabular}{|rccr|} \hline  \hline
Quantity & Experimental Value & SM &
Correlation \\ \hline

$\epsilon_L(u)$ & $0.332 \pm 0.016$ & $0.345 \pm 0.001$ & \  \\
$\epsilon_L(d)$ & $-0.438 \pm 0.012$ & $-0.429 \pm 0.001$ & non - \\
$\epsilon_R(u)$ & $-0.178 \pm 0.013$ & $-0.156 $ & Gaussian \\
$\epsilon_R(d)$ & $-0.026^{+0.075}_{-0.048}$ & 0.078  & \  \\
\hline
$g^2_L$ & $0.3017 \pm 0.0033$ & 0.303 $ \pm 0.001$ & \ \\
$g^2_R$ & $0.0326 \pm 0.0033$ & 0.030  & small \\
$\theta_L$ & $2.50 \pm 0.035$ & 2.46 & \ \\
$\theta_R$ & $4.58^{+0.46}_{-0.28}$ & 5.18  & \ \\ \hline
$g^{\nu e}_A$ & $-0.507 \pm 0.014 $ & $ -0.506 \pm 0.001$ & \ \ \ $-0.04$ \\
$g^{\nu e}_V$ & $-0.041 \pm 0.015$ & $ -0.037 \pm 0.001$ & \ \\ \hline
$C_{1u}$ & $-0.214 \pm 0.046$ & $-0.189 \pm 0.001$ & $-0.995 \;\; -0.79$
\\
$C_{1d}$ & $0.359 \pm 0.041$ & $0.341 \pm 0.001$ & $ \;\;\; 0.79$ \\
$C_{2u} - \frac{1}{2}C_{2d}$ & $ -0.04 \pm 0.13$ & $-0.051 \pm 0.002$ &
\ \ \\ \hline \hline
\end{tabular}
\caption[]{Values of the model-independent neutral current parameters,
compared with the standard model predictions (SM)
using $M_Z = 91.1888 \pm 0.0044$ GeV and $m_t = 175 \pm 11$ GeV
for $M_H$ = 300 GeV.
$g^2_{L,R}$ are
defined by $g^2_{L,R} = \epsilon_{L,R}(u)^2 + \epsilon_{L,R}(d)^2$,
while $\theta_i, \; i = L$ or $R$, is defined as $\tan^{-1} \left[
\epsilon_i (u)/\epsilon_i (d) \right]$.}
\label{tab11}
\end{table}

There have been a number of neutrino-electron
experiments~\cite{nuq,a5}, as described in the article by Panman.
The best measured are $\nu_\mu e \ra \nu_\mu e$ and $\bar{\nu}_\mu e
\ra \bar{\nu}_\mu e$.  The most general \lag\ allowed by a gauge
theory with family universality and left-handed neutrinos is
\beq -L^{\nu e} = \frac{G_F}{\sqrt{2}} \bar{\nu}
\gamma^\mu (1 - \gamma^5) \nu J^e_\mu, \label{eq327.1} \eeq
where
\beq J^e_\mu = \bar{e} \gamma_\mu
\left[ g^{\nu e}_V - g_A^{\nu e} \gamma^5
\right] e, \label{eqch123} \eeq
and $g^{\nu e}_V$ and $g^{\nu e}_A$ refer to
the vector an axial vector couplings
of the electron in the four-fermi interaction.  The $\nu_\mu$
reactions determine $g^{\nu e}_V$ and $g^{\nu e}_A$ up to a four-fold
ambiguity, as
is shown in Figure~\ref{fig46}.  One of the solutions corresponds to
the standard model and the others to the interchange of $g^{\nu e}_V$ with
$g^{\nu e}_A$ and an overall sign change.

Some of the ambiguity can be eliminated by considering the additional
reactions $\nu_ee \ra \nu_ee$ and $\bar{\nu}_ee \ra \bar{\nu}_e e$,
for which both charged and neutral currents contribute and interfere.
%
\begin{figure}
\small \def\baselinestretch{1} \normalsize
\postbb{70 230 530 670}{/home/pgl/fort/nc/graph/modind/xxnue.ps}{0.6}
\caption[]{Allowed regions in the $g^{\nu e}_V$, $g^{\nu e}_A$ plane from
neutrino-electron data, compared with the standard model prediction
as a function of $s^2_W$.}
\label{fig46}
\end{figure}
%
%
The $\nu_ee$ reaction has been measured at LANL~\cite{n4}; it yields
the additional constraint shown in Figure~\ref{fig46}.  Finally, there
is  the Savannah River
reactor $\bar{\nu}_{e}e$ experiment~\cite{Reines}, which yields a
different allowed contour.  From these one sees that the $\nu_e$ data
determine the couplings up to a two-fold ambiguity, one of which
(the axial-vector dominant)
corresponds to the standard model.  It is possible to eliminate
the second (vector dominant)
solution  by simultaneously considering $e^+e^- \ra \mu^+ \mu^-$
data if one assumes that the amplitude factorizes into neutrino and
charged-lepton factors.  This is true if the neutral current amplitude
is dominated by the exchange of a single $Z$ boson.  That is certainly
a reasonable assumption for gross purposes such as eliminating the
vector-dominant solution.  However, it should be warned that many
types of new physics, such as extra $Z'$ bosons, break the
factorization, and care should be applied when relating the two types of
reactions.  The expressions for
$g^{\nu e}_V$ and $g^{\nu e}_A$ in the standard
model are shown in Table~\ref{tab10}, and the numerical values
extracted from the data are compared with the standard model
predictions in Table~\ref{tab11}.

There have been a number of measurements of the interference between
weak and electromagnetic amplitudes in the electron-quark system.


The parity-violating eq interaction generated by $Z$ exchange can be
expressed in an arbitrary gauge theory by
\beq + L^{eH} = \frac{G_F}{\sqrt{2}} \sum_{i = u,d}
\left[ C_{1i} \bar{e} \gamma^\mu \gamma^5 e \bar{q}_i \gamma_\mu q_i
 +  C_{2i} \bar{e} \gamma^\mu e \bar{q}_i \gamma_\mu \gamma^5
q_i \right], \label{eqch124} \eeq
where the $C_{1i}$ represent axial electron and vector quark currents
while the $C_{2i}$ represent vector electron and axial quark currents.
There are additional parity-conserving pieces, which are negligibly
small compared to electromagnetism.  There have been several
experiments on the electron-quark coupling.  Most notably, the
polarized electron-deuteron asymmetries $e^{(\uparrow\!\downarrow)} D
\ra e X$ from SLAC~\cite{SLAC}, and, more recently, other experiments,
such as an $e^{(\uparrow \! \downarrow)} Be \ra eBe^* $ asymmetry
experiment from Mainz~\cite{n5}, $e^{(\uparrow \downarrow)}C$ from
Bates~\cite{bates}, and the BCDMS $\mu C$  asymmetry experiment~\cite{BCDMS}
done at CERN. These are described in detail in the article by Souder.

In addition, parity violation manifests itself in atomic physics by
leading to parity-violating mixtures between $S$ and $P$ wave states.
The subject has had a long and difficult history, but in recent years
the experiments have advanced greatly; in particular, there are now
high precision measurements of parity violation in the cesium atom,
first in Paris~\cite{Paris} and more recently in
Boulder~\cite{v4}, as described in the article by
Masterson and Wieman.
Cesium is a very clean atom to study
theoretically \cite{v5}: it is a single electron outside of a tightly-bound
core.  Recent calculations of the matrix elements needed to interpret
the experimental results are accurate at the 1\% level\footnote{
In the future it may be possible to eliminate most theoretical
uncertainties by comparing the parity-violating effects in
different isotopes of the same atom.},
as described in detail by Blundell, Johnson, and Sapirstein.
The experimental precision should equal this soon.  The $C_{1i}$ couplings
are much better determined than the $C_{2i}$ because the $C_{1i}$
operators are coherent with respect to the nucleons in a heavy
nucleus, while the $C_{2i}$ operators couple to nucleon spin. Also, the
$C_{2i}$ operators involve theoretical ambiguities from the $s$ quark
content of the nucleon~\cite{240a}
and nuclear anapole moment effects~\cite{240a}.  The
experimental constraints are compared with the standard model
predictions in Figure~\ref{fig49} and Table~\ref{tab11}.  The
agreement is excellent.  It is apparent that the cesium results are
especially useful for constraining types of new physics which shift
the predicted parameters in a direction perpendicular to the narrow
band in Figure~\ref{fig49}~\cite{r10,a30}.

\begin{figure}
\small \def\baselinestretch{1} \normalsize
\postbb{70 230 530 670}{/home/pgl/fort/nc/graph/modind/xxeq1.ps}{0.6}
\caption[]{Constraints from $eD$ scattering and atomic parity violation,
compared with the standard model prediction
as a function of $s^2_W$.}
\label{fig49}
\end{figure}

Weak-electromagnetic interference~\cite{epem} can also be observed in
$e^+e^-$ annihilation into $\mu^+ \mu^-$, $\tau^+ \tau^-$, $b\bar{b}$,
$c \bar{c}, \cdots$.  
Experiments were done at SLAC, DESY and TRISTAN
well below the $Z$-pole, where the dominant contribution is photon
exchange, with the $Z$ a perturbation.  The interference can lead to a
forward-backward asymmetry, $A_{FB}$,
in the direction of the $\mu^-$ with respect to the $e^-$.
Well below the $Z$-pole the
asymmetry is predicted in the standard model to be
\beq A_{FB} = \frac{- 3G_F}{16 \sqrt{2} \pi \alpha} \; \frac{s
M_Z^2}{M_Z^2 - s       }, \label{eqch125} \eeq
for $\sqrt{s} \ll M_Z$, where $\sqrt{s}$ is the total center of mass
energy.  This depends only on the axial couplings of the electron and
muon, and is therefore independent of $\sin^2\theta_W$ except for a
(small) indirect dependence via $M_Z$.  Therefore, it is essentially
an absolute prediction of the model\footnote{One can extract $\sinn$ by
using the tree-level relation $G_F/\sqrt{2} \pi \alpha =
(2 \cos^2 \theta_W \sinn M_Z^2)^{-1}$, and
combining  $A_{\rm FB}$ with the value of
$M_Z$ measured in other experiments.  In fact, the asymmetry
limits the deviation from the standard model, while $M_Z$
yields $\sinn$.}.
The magnitude of the
asymmetry is predicted to
increase approximately linearly with $s$.  The data is compared with
the prediction in Figure~\ref{fig51}; the agreement is excellent.
The annihilation into hadrons is discussed in detail in the article
by Haidt.


The model independent analyses established that most of the four-fermi
interactions are uniquely determined, consistent with the standard
model and in disagreement with many competing gauge theories.
The more precise $Z$-pole measurements from LEP and SLC
subsequently excluded large classes of small deviations from
the standard model predictions. However, there are many types
of new physics which do not directly affect the $Z$ or its
couplings and to which the $Z$-pole experiments are blind.
The low energy experiments therefore still play a significant role
in probing for such types of new physics.

\begin{figure}
\small \def\baselinestretch{1} \normalsize
\postbb{50 230 530 690}{/home/pgl/fort/nc/graph/modind/xxee.ps}{0.6}
\caption[]{Forward-backward asymmetry for $e^+e^- \ra \mu^+\mu^-$
as a function of the square of the center of mass
energy, compared with the prediction of the standard model (which is
almost independent of $\sin^2\theta_W$).}
\label{fig51}
\end{figure}

\vglue 0.6cm
\section{\elevenbf Beyond the Standard Model}
\subsection{\elevenit Unification or Compositeness}
\vglue 0.4cm

Most work in particle physics today is directed towards searching for the
new physics beyond the standard model.  Although there are many theoretical
ideas for the nature of such new physics most possibilities fall into one
of two general categories.

The first, which I describe as the Bang scenario, involves the unification
of the interactions.  In such schemes there is generally a grand desert up
to a grand unification (GUT) or Planck scale ($M_P$).  This is the natural
domain of elementary Higgs fields, supersymmetry, GUTs, and superstring
theories.  If nature should choose this route there is a possibility of
probing to $M_P$ and to the very early universe.  There are hints from
coupling constant unification that this may be the correct path.  Some of
the implications are that there should be supersymmetry, which can
ultimately be probed by finding the new superpartners at the LHC.
Secondly, one expects to have a light Higgs boson, which acts much like the
standard model Higgs except that it must be lighter than $110 - 150$~GeV,
which should be detectable at the LHC or possibly at LEP 2.  (The
standard model Higgs could be as heavy as 600 $-$ 1000~GeV.)  Finally, a
very important prediction of at least the simplest cases is that one
expects an {\it absence} of deviations from the standard model predictions
for precision electroweak tests, CP violation, or rare $K$ decays, because
of the decoupling of the heavy superpartners.  Of course, it is hard to
take the observed absence of such deviations as compelling evidence for
supersymmetric unification, but they are nevertheless suggestive.  Some
such schemes also lead to predictions for $m_b$, proton decay, neutrino
masses, and rare decays.

If the coupling constant unification is not just an accident there are very
few types of new  physics other than supersymmetry that could be present
without spoiling it (unless two new effects cancel).  These include
additional heavy $Z'$ bosons, gauge singlets, and a small number of
new sequential, mirror, or exotic fermion families.

The other general possibility is the Whimper scenario, in which nature
consists of onion-like layers of matter at shorter and shorter distance
scales.  This is the domain of composite fermions and scalars and of
dynamical symmetry breaking.  Experimental limits imply that
any new layer of compositeness
would have to be strong binding, and is therefore not analogous to previously
observed levels.  If nature should choose
this route, then at most one more layer would be accessible to us at the
LHC and future colliders.  Such schemes generally predict significant rates
for rare decays such as $K \ra \mu e$.  This is a generic feature of almost
all such models, and the fact that they have not been observed is a severe
problem for the general approach and has made it difficult to construct
realistic models.  If one somehow evades the problem of rare decays one
still generally expects to see significant effects in
precision observables, including new 4-fermi operators,
decrease of the $Z \ra b\bar{b}$ partial width,
and modifications
to $\rho_0$ and to the parameters $S$, $T$, and $U$.
The fact that these have not been seen constitutes an additional serious
difficulty for most such models.  In the future one would also expect to
see new particles and anomalous interactions among gauge bosons.

\vglue 0.6cm
\subsection{\elevenit Searches for New Physics}
\vglue 0.4cm
Since there is no evidence for deviations from the standard model, the
boson and neutral current data (and also precision charged current data)
can be used to set limits on many kinds
of possible new physics.
These include (a)
heavy $Z'$ bosons \cite{n2a}-\cite{n2d};
(b) the $\rho_0$ parameter, associated with higher-dimensional Higgs
representations or other new sources of $SU_2$-breaking \cite{rhonot};
and (c) classes  of new physics (such as technicolor or new
multiplets of fermions or scalars) which only affect the $Z$, $W$,
and neutral current observables via gauge self-energy
diagrams \cite{a35}-\cite{a35e}.
Other applications include (a) verifying
the canonical (\ie left-handed doublets, right-handed
singlets) weak isospin assignments of the known quarks and leptons
\cite{v15,v16}; (b) searches for mixing between ordinary and exotic
fermions (\eg left-handed singlets or right-handed doublets,
which are predicted in many extensions of the
standard model) \cite{v35,LLM};
(c) searches for leptoquarks or new
four-fermi operators associated with compositeness \cite{LLM,r10,a30};
(d) searching for anomalous contributions to the $Z b \bar{b}$ vertex
\cite{bb1},
such as may be generated by light superpartners \cite{bbbarref}
or extended technicolor interactions \cite{bb2}-\cite{bb2a}.
The sensitivity of existing and projected experiments for various classes
of new physics are described in detail in \cite{LLM}.

\vglue 0.6cm
\subsection{\elevenit Supersymmetry and Precision Experiments}
\vglue 0.4cm

Let us now consider how the predictions for the precision observables are
modified in the presence of supersymmetry.  There are basically three
implications for the precision results.  The first, and most important, is
in the Higgs sector.  In the standard model the Higgs mass is arbitrary.  It
is controlled by an arbitrary quartic Higgs coupling, so that $M_H$ could
be as small as 60 GeV (the experimental limit) or as heavy as a TeV.  The
upper bound is not rigorous: larger values of $M_H$ would correspond to
such large quartic couplings that perturbation theory would break down.
This cannot be excluded, but would lead to a theory that is qualitatively
different from the (perturbative) standard model.  In particular, there are
fairly convincing triviality arguments, related to the running of the
quartic coupling, which exclude a Higgs which acts like a distinct elementary
particle for $M_H$ above $O(600$~GeV) \cite{a25}.

However, in supersymmetric extensions of the standard model the quartic
coupling is no longer a free parameter.  It is given by the squares of
gauge couplings, with the result that all supersymmetric models have at
least one Higgs scalar that is relatively light, typically with a mass
similar to the $Z$ mass.  In the minimal supersymmetric standard model
(MSSM) one has $M_H < 150$~GeV\footnote{At tree-level,
$M_H < M_Z$.}, which generally acts just like the standard
model Higgs\footnote{This is true if the second Higgs doublet is much
heavier than $M_Z$.} except that it is necessarily light.

In the standard model there is a strong $m_t - M_H$ correlation, and one
has the prediction for $m_t$ is (\ref{mtmh}).
For the standard model range $60< M_H < 1000$~GeV this corresponds to
$ m_t = 175 \pm11 ^{+17}_{-19}$ GeV (SM).
However, in MSSM one has the smaller range $60 < M_H < 150$~GeV, leading to
the lower prediction
\beq m_t = 160^{+11 +6}_{-12 -5}  \ (\rm MSSM).\eeq
This is on the low side of the CDF range, $(174 \pm 16$~GeV), but is
certainly consistent.
Similarly, the indirect data predict
\beq \hat{s}^2_Z = 0.2316(3)(1) \ (\rm MSSM) \label{smssm}  \eeq
and
\beq \alpha_s = 0.126(5)(1) \ (\rm MSSM), \eeq
which differ slightly from the standard model values in Table~\ref{tab4}
because of the lower $M_H$ range.

There can be additional effects on the radiative corrections due to
sparticles and the second Higgs doublet that must be present in the
MSSM \cite{susyrad}.
However, for most of the allowed parameter space one has $M_{\rm new}
\gg M_Z$, and
the effects are negligible by the decoupling theorem.  For example, a large
$\tilde{t} - \tilde{b}$ splitting would contribute to the $\rho_0$
($SU_2$-breaking) parameter to be discussed below, leading to a smaller
prediction for $m_t$, but these effects are negligible for $m_{\tilde{q}}
\gg M_Z$.  Similarly, there would be new contributions to the $Z\ra
b\bar{b}$ vertex for $m_{\chi^\pm}$, $m_{\tilde{t}}$, or $M_H^\pm \sim M_Z$.

There are only small windows of allowed parameter space for which the new
particles contribute significantly to the  radiative corrections.  Except
for these, the only implications of supersymmetry from the precision
observables are: (a) there is a light standard model-like Higgs, which in
turn favors a smaller value of $m_t$.  Of course, if a light Higgs were
observed it would be consistent with supersymmetry but would not by itself
establish it.  That would require the direct discovery of the
superpartners, probably at the LHC.  (b) Another important implication of
supersymmetry, at least in the minimal model, is the {\it absence} of other
deviations from the standard model predictions.  (c) In supersymmetric
grand unification one expects the gauge coupling constants to unify when
extrapolated from their low energy values \cite{recent}.  This is consistent
with the data in the MSSM but not in the ordinary standard model (unless
other new particles are added).  This is not actually a modification of the
precision experiments, but a prediction for the observed gauge couplings.
Of course, one could have supersymmetry without grand unification.

\vglue 0.6cm
\subsection{\elevenit (Supersymmetric) Grand Unification}
\vglue 0.4cm
\label{gutsec}
It is interesting to compare the value of $\hat{s}^2_Z$
in Table~\ref{tab4} or eqn.~(\ref{smssm})
with the only models available which predict it, namely grand unified
theories \cite{recent}-\cite{r3}.

In a grand unified theory there is only one underlying gauge
coupling constant, and when the low energy couplings are
extrapolated to high energy they are expected to (approximately)
meet at the unification scale $M_X$ above which symmetry
breaking can be neglected.
We define the couplings $g_s = g_3$, $g=g_2$, and $g' =
\sqrt{3/5} g_1$ of the standard model $SU_3 \x SU_2 \x U_1$ group, and
the fine-structure constants $\alpha_i = g^2_i/4\pi$.  The extra
factor in the definition of $g_1$ is a normalization
condition~\cite{r3}.  The couplings are expected to meet only if the
corresponding group generators are normalized in the same way.
However, the standard model generators are conventionally normalized
as $\Tr (Q^2_s) = \Tr (Q^2_2) = 5/3 \Tr (Y/2)^2$, so the factor
$\sqrt{3/5}$ is needed to compensate. Thus,
\beq \sin^2 \theta_W = \frac{g'^{2}}{g^2 + g'^{2}} = \frac{g^2_1}{
\frac{5}{3} g^2_2 + g_1^2} \stacksub{\ra}{g_1 = g_2}
\frac{3}{8}.\eeq
One expects $\sin^2\theta_W = 3/8$ at the unification scale~\cite{r3}
for which $g_1 = g_2$.

To test the unification, one starts with the couplings at \mz, which
are now very well known from the LEP and low energy data.
Using  as inputs
$\alpha^{-1}(M_Z) =127.9 \pm 0.1$  \cite{a13},\
 \shz \ $= 0.2316\pm 0.0003$,
 and\footnote{\als is considerably less well-determined than
$\alpha(M_Z)$ and \shz. I will therefore usually take \als as a
prediction rather than an input.
When it is used as an input, however, I will use the conservative
range $0.12 \pm 0.01$.}
$ \alpha_s (M_Z) = 0.12 \pm 0.01$, \
 \ one obtains
 \begin{eqnarray}
 \alpha^{-1}_1(M_Z)
 & \equiv & \frac{3}{5} \alpha^{-1}(M_Z) \hat{c}^2_Z
  = 58.97 \pm 0.05 \nonumber \\
 \alpha ^{-1}_2 (M_Z)& \equiv & \alpha^{-1}(M_Z) \
 \shz \ = 29.62 \pm 0.04 \nonumber \\
 \alpha^{-1}_3 (M_Z)& \equiv & \alpha^{-1}_s(M_Z) = 8.3 \pm 0.7.
 \end{eqnarray}

\begin{figure}
\small \def\baselinestretch{1} \normalsize
\postbb{50 70 530 680}{/u/pgl/fort/nc/graph/gut/xxgut.ps}{.8}
\caption[]{Running couplings in (a) the standard model (SM) and (b) in the
minimal supersymmetric extension of the standard model (MSSM) with two
Higgs doublets for
$M_{SUSY} = M_Z$.
The corresponding figure for $M_{SUSY} = 1$ TeV is
almost identical. It is
seen that the couplings unify at $\simeq 10^{16}$~GeV in the MSSM.
The effects of threshold uncertainties are seen in Figures
\ref{gutb} and \ref{gutc}.}
\label{guta}
\end{figure}

These may be extrapolated to high energy using
the two-loop renormalization group equations
\beq \frac{d\alpha^{-1}_i}{d \ln \mu} = - \frac{b_i}{2 \pi} -
\sum^3_{j=1} \frac{b_{ij} \alpha_j}{8\pi^2} . \label{twoloop} \eeq
The 1-loop coefficients are
\beq b_i = \left(\begin{array}{c} 0 \\ -\frac{22}{3} \\ - 11
\end{array} \right) + F \left(\begin{array}{c} \frac{4}{3} \\
\frac{4}{3} \\ \frac{4}{3} \end{array} \right) + N_H
\left(\begin{array}{c} \frac{1}{10} \\ \frac{1}{6} \\ 0 \end{array}
\right) ,\eeq
assuming the standard model. $F$ is the number of fermion families and
$N_H$ is the number of Higgs doublets.  In the MSSM,
\beq b_i = \left(\begin{array}{c} 0 \\ -6 \\ - 9 \end{array} \right) +
F \left(\begin{array}{c} 2 \\ 2 \\ 2 \end{array} \right) + N_H
\left(\begin{array}{c} \frac{3}{10} \\ \frac{1}{2} \\ 0 \end{array}
\right) ,\eeq
where the difference is due to the additional particles
in the loops.  The 2-loop coefficients can be found in
\cite{v31}.
Equation \ref{twoloop}
can be integrated to yield
\beq  \alpha^{-1}_i (\mu) = \alpha_i^{-1} (M_Z) - \frac{b_i}{2\pi}
\ln \left( \frac{\mu}{M_Z} \right)
   + \sum^3_{j=1} \frac{b_{ij}}{4\pi b_j} \ln \left[
   \frac{\alpha^{-1}_j (\mu)}{\alpha_j^{-1} (M_Z)} \right] ,\eeq
   for an arbitrary scale $\mu$.
   To first approximation one can neglect
   the last (2-loop) term, in which case the inverse coupling constant
   varies linearly with $\ln \mu$.
   However, the 2-loop terms must be kept in the final analysis.
In a grand unified theory one expects that the three couplings will
approximately meet at $M_X$,
\beq \alpha^{-1}_i (M_X) = \alpha_G^{-1} (M_X) - \Delta_i.
\eeq
The $\Delta_i$ are small corrections \cite{r9}
associated with the low energy
threshold (\ie \mt \ and the new sparticles and Higgs not degenerate
with \mz), the high scale thresholds
($m_{\rm heavy}
\neq M_X$), or with non-renormalizable operators.

The running couplings in the standard model are shown in Figure
\ref{guta}a, ignoring threshold corrections.  They clearly
do not meet at a point, thus ruling out simple
grand unified theories such as $SU_5, SO_{10}$, or $E_6$ which break in
a single step to the standard model \cite{v30}. Of course, such models
are also excluded by the non-observation of proton decay,
but this independent
evidence is welcome.
\begin{figure}
\small \def\baselinestretch{1} \normalsize
\postbb{60 220 530 690}{/u/pgl/fort/nc/graph/gut/xxszgut.ps}{.6}
\caption[]{90\% CL regions in \shz \ vs $m_t$,
 compared with the predictions of
ordinary and SUSY--GUTs. The smaller ranges of uncertainties
are from $\alpha (M_Z)$ and $\alpha_s (M_Z)$ only, while the
larger range includes the various low and high scale uncertainties,
added in quadrature. The predictions for degenerate SUSY masses
at \mz \ and 1 TeV are shown for comparison. Updated from \cite{a22aa}.}
\label{gutb}
\end{figure}

On the other hand, in the minimal supersymmetric extension of the
standard model the couplings do meet within the experimental
uncertainties \cite{v31,amaldi,recent}.
This is illustrated in Figure \ref{guta}b
for the case in which
all of the new particles have a common mass $M_{SUSY} = M_Z$. Almost
identical curves are obtained for larger $M_{SUSY}$, such as 1 TeV.
(In practice, the splittings between the sparticle masses
are more important than the average value \cite{rr,r9}.)
The unification scale $M_X$ is sufficiently large $(>10^{16}$ GeV) that
proton decay by dimension$-6$ operators is adequately suppressed,
although there may still be a problem with dimension$-5$ operators
\cite{v33}. This success is encouraging for supersymmetric grand unified
theories such as $SUSY$--$SU_5$ or $SUSY$--$SO_{10}$.

To display the theoretical uncertainties, it is conventional to
use $\alpha (M_Z)$ and $\alpha_s (M_Z)$ to predict \shz.
Using
$\alpha^{-1}(M_Z) = 127.9 \pm 0.1$ and \alsz = 0.12 $\pm$ 0.01
one predicts
\beqa \shz & = & 0.2334\pm
0.0025 \pm 0.0025\ {\rm (MSSM)}, \nonumber \\
\shz & = & 0.2100\pm 0.0025 \pm
0.0007 \ {\rm (SM)}, \label{eqgut} \eeqa
where the first uncertaintly is from $\alpha_s$ and $\alpha^{-1}$,
and the second is an estimate of theoretical uncertainties
from $m_t$, the superspectrum, high-scale thresholds, and possible
non-renormalizable operators \cite{a22aa}.
The MSSM prediction is in agreement with the
experimental value $0.2316(3)(1)$, while the SM prediction is
in conflict with the data.
These results are displayed in Figure \ref{gutb}.

Because of the large uncertainty in \alsz, it is convenient
to invert the logic and use the precisely known
$\alpha^{-1}$ and \shz \
to predict \alsz:
\beqa \alsz & = & 0.129 \pm 0.002 \pm 0.008 \ {\rm (MSSM)}, \nonumber \\
\alsz & = & 0.073 \pm 0.001 \pm 0.001 \ {\rm (SM)}, \label{eqgut1} \eeqa
where again the second error is theoretical.
 It is seen that the
 SUSY prediction is in agreement with the experimental
 \alsz = 0.12 $\pm$ 0.01, while the
 simplest ordinary GUTs are excluded.
 The central value
 prefers the larger values of \alsz \  suggested
 by the $Z$-pole data over some of the low energy
 determinations in Table~\ref{tab5}, but the theoretical uncertainties are
 comparable to the error on the observed \alsz \ (which is also
 dominated by theory).
 If the low energy values turn out to be true, supersymmetric
 unification would require large but not unreasonable threshold
 corrections.
 The \alsz \ predictions are shown in Figure \ref{gutc}.
\begin{figure}
\small \def\baselinestretch{1} \normalsize
\postbb{65 75 525 680}{/home/pgl/fort/nc/graph/gut/xxagut.ps}{0.8}
\caption[]{Predictions for \alsz \ from $\alpha^{-1}$ and
\shz \ in ordinary and SUSY GUTs.
The dashed lines represent the experimental range
0.12 $\pm$ 0.01. In the SUSY (MSSM) case the error bar
includes the theoretical uncertainties, added in quadrature.
The smaller error bars are for degenerate sparticle masses
at \mz \ and 1 TeV. Updated from \cite{a22aa}.}
\label{gutc}
\end{figure}

 The success of the coupling constant unification, which is insensitive
 to the gauge group and the number of complete families, provides a
 hint that supersymmetric grand unification (or some superstring
 imitator) may be on the right track.
 Of course it is possible that the success may be an accident.
 Similarly, there are many more complicated schemes which could
yield coupling constant unification, such as those involving
a large group breaking in two or more stages to the standard model,
or those with ad hoc new representations split into light
and heavy components. However, the MSSM is the only scheme in
which  the unification is a prediction rather than being achieved by
adjusting new parameters or representations.
 Perhaps the coupling constants may
 indeed prove to be the ``first harbinger of supersymmetry''
 \cite{amaldi}.

Unless the apparent coupling constant unification is an accident,
there are stringent restrictions on the types of new ``SUSY-safe''
physics which do not drastically disturb the predictions (unless,
of course, one allows two large effects to cancel). These
are: supersymmetry (required), additional heavy $Z'$ gauge bosons
or additional gauge groups which commute with the standard model,
additional complete ordinary or mirror fermion families
and their superpartners (the
neutrinos would have to be very heavy because of the LEP result
$N_\nu = 2.988 \pm 0.023$), complete exotic fermion
supermultiplets  such as occur in some $E_6$ models, or gauge singlets.

There are many other implications of supersymmetric grand unification.
These include proton decay \cite{v33};
Yukawa unification (the prediction of $m_b/m_\tau$),
which leads to stringent constraints on the ratio $\tan \beta$ of
the vacuum expectation values of the two Higgs doublets which give
mass to the $t$ and $b$, respectively \cite{v33a};
the upper limit on the standard model-like Higgs \cite{v33b,v33bb};
cold dark matter \cite{v33c}; neutrino mass \cite{v33d}; and
a possible connection with superstring theories~\cite{v33e}.

\vglue 0.6cm
\subsection{\elevenit Extended Technicolor/Compositeness}
\vglue 0.4cm

In contrast to unification/supersymmetry,
the other major class  of extensions, which includes
compositeness and dynamical symmetry breaking, leads to many implications
at low energies.  The most important are large flavor changing neutral
currents (FCNC).  Even if these are somehow evaded one generally expects
anomalous contributions to the $Z \ra b\bar{b}$ vertex, typically
$\Gamma(b\bar{b}) < \Gamma^{SM}(b\bar{b})$ in the simplest extended
technicolor (ETC) models \cite{bb2}.  Similarly, one expects $\rho_0 \neq
1$, and $S_{\rm new} \neq 0, T_{\rm new} \neq 0$, where $\rho_0$, $S_{\rm
new}$, and $T_{\rm new}$ parameterize certain types of new physics, as will
be described below.  Finally, in theories with composite fermions one
generally expects new 4-fermi operators generated by constituent
interchange, leading to effective interactions of the form
\beq L = \pm \frac{4 \pi}{\Lambda^2} \bar{f}_1 \Gamma f_2 \bar{f}_3
\hat{\Gamma}f_4.\eeq
Generally, the $Z$-pole observables are not sensitive to such operators,
since they only measure the properties of the $Z$ and its
couplings.
However, low energy experiments are sensitive.  In particular, FCNC
constraints typically set limits of order $\Lambda \geq O(100\; {\rm TeV})$
on the scale of the operators unless the flavor-changing effects are
fine-tuned away.  Even then there are significant limits from other flavor
conserving observables.  For example, atomic parity violation \cite{v4} is
sensitive to operators such as \cite{r10,LLM}
\beq L = \pm \frac{4\pi}{\Lambda^2} \bar{e}_L \gamma_\mu e_L
\bar{q}_L \gamma^\mu q_L .\eeq
The existing data already sets limits $\Lambda > O(10$~TeV), as
described in section~\ref{fourferm}.  Future
experiments should be sensitive to $\sim$ 40~TeV.

\vglue 0.6cm
\subsection{\elevenit The $Zb\bar{b}$ Vertex}
\vglue 0.4cm

\label{zbbsec}
The $Zb\bar{b}$ vertex is especially interesting, both in the standard model
and in the presence of new physics.  In the standard model there are
special vertex contributions, shown in Figure~\ref{fig3},
which depend quadratically on the top quark
mass. Their value is shown approximately in (\ref{eq26a}). $\Gamma (b\bar{b})$
actually decreases with $m_t$, as opposed to other widths which all
increase due to the $\hat{\rho}$ parameter.  The $m_t$ and $M_H$
dependences in $\hat{\rho}$ are strongly correlated, but the special vertex
corrections to $\Gamma(b\bar{b})$ are independent of $M_H$, allowing a
separation of $m_t$ and $M_H$ effects.

\begin{figure}[htb]
\postbb{85 30 500 740}{/home/pgl/fort/nc/graph/misc/zbb.ps}{0.5}
\caption[]{(a) Standard model vertex corrections to $Z \ra b\bar{b}$. (b) New
contributions in supersymmetry. (c) New contributions in extended
technicolor.}
\label{fig3}
\end{figure}

The vertex is also sensitive to a number of types of new physics.
One can parameterize such effects by \cite{bb1}
\beq \Gamma {(b\bar{b})} \ra \Gamma^{SM} {(b\bar{b})} \left( 1 +
\delta^{\rm new}_{bb} \right) \sim \Gamma^0 {(b\bar{b})} \left( 1
+ \delta^{\rm SM}_{bb} + \delta^{\rm new}_{bb} \right).
\eeq
If the new physics gives similar contributions to vector and axial vector
vertices then the effects on $A_{\rm FB}^{b}$ are negligible.  In
supersymmetry one can have both positive and negative contributions
\cite{bbbarref}.  In particular, light $\tilde{t} - \chi^{\pm}$ can give
$\delta^{\rm SUSY}_{bb} > 0$, as is suggested by the data, while light
charged Higgs particles yield $\delta^{\rm Higgs}_{bb} < 0$.  In
practice, both effects are too small to be important in most allowed
regions of parameter space~\cite{aa1}.
In extended technicolor (ETC) models there are
typically new vertex contributions generated by the same ETC interactions
that are needed to generate the large top quark mass.  It has
been argued that these are typically large and negative \cite{bb2},
\beq \delta^{\rm ETC}_{bb} \sim - 0.056 \xi^2 \left(
\frac{m_t}{150{\rm GeV}} \right),\eeq
where $ \xi$ is a model dependent parameter of order unity.
They may be smaller in models with walking technicolor, but nevertheless
are expected to be negative and significant \cite{bb2b}.  This is in
contrast to the data, which suggests a positive contribution if any,
implying a serious problem for many ETC models.  One possible way out are
models in which the ETC and electroweak groups do not commute, for which
either sign is possible \cite{bb2a}.

Another possibility is mixing  between the $b$ and exotic heavy fermions
with non-canonical weak interaction quantum numbers.  Many extensions of
the standard model predict, for example, the existence of a heavy
$D_L$, $D_R$, which are both $SU_2$ singlet quarks with charge $-1/3$.
These can mix with the $d$,
$s$, or $b$ quarks, but one typically expects such mixing to be largest for
the third generation.  However, this mechanism gives a negative
contribution
\beq \delta_{bb}^{D_L} \sim -2.3 s^2_L \eeq
to $\delta^{\rm new}_{bb}$, where $s_L$ is the sine of the $b_l - D_L$
mixing angle. $R_b$ can be increased if there is an additional
heavy $Z'$ boson which only couples to the third family~\cite{nuzp}.

One can extract $\delta^{\rm new}_{bb}$ from the data, in a global fit to
the standard model parameters as well as $\delta^{\rm new}_{bb}$.  This
yields~\cite{a7c}
\beq \delta^{\rm new}_{bb} = 0.023 \pm 0.011 \pm 0.003,  \eeq
which is $\sim 2 \sigma$ above zero.  This value is hardly changed when
one allows additional new physics, such as described by the $S$, $T$, and
$U$ parameters.  $\delta^{\rm new}_{bb}$ is correlated with
$\alpha_s(M_Z)$, because
one can describe $R = \Gamma ({\rm had})/\Gamma(\ell
\bar{\ell})$ with a smaller QCD correction to
$\Gamma(\rm had)$~\cite{a7c,nuzp}.
Allowing for $\delta^{\rm new}_{bb}$,
one obtains $\alpha_s(M_Z) = 0.111 \pm 0.009 \pm 0$,
considerably smaller than the standard model value $0.127 (5)(2)$. Allowing
$\delta_{bb}^{\rm new} \neq 0$ has negligible effect on $\hat{s}^2_Z$ or
$m_t$. One can also perform more detailed fits allowing
separate corrections to the left and right-handed $b$
couplings~\cite{a7c,blbr}.
Using both $R_b$ and $A^{0b}_{FB}$ (as well as all of the other data)
as constraints, one finds that the anomaly should be in the $b_R$
coupling, as can be seen in equation (\ref{biso}).

\vglue 0.6cm
\subsection{\elevenit $\rho_0$: Nonstandard Higgs  or Non-degenerate Heavy
Multiplets}
\vglue 0.4cm

One parameterization of certain new types of physics is the parameter
\beq \rho_0 \equiv \frac{M^2_W}{M^2_Z \hat{c}^2_Z \hat{\rho}}
 \label{rhodef}.  \eeq
$\rho_0$ is exactly unity in the standard model, and any
deviation would indicate
new sources of $SU_2$ breaking
other than the ordinary Higgs doublets or the top/bottom splitting.
New physics can affect $\rho_0$ at either the tree or loop-level
\beq \rho_0 = \rho_0^{\rm tree} + \rho_0^{\rm loop}.\eeq
The tree-level contribution is given by Higgs representations
larger than doublets, namely,
\beq \rho_0^{\rm tree} = \frac{\sum_i \left( t^2_i -  t_{3i}^2
+ t_i \right) |\langle \phi_i \rangle|^2}{ \sum_i 2 t_{3i}^2
|\langle \phi_i \rangle|^2},  \label{eqerica}\eeq
where $t_i$ ($t_{3i}$) is the weak isospin (third component) of the
neutral Higgs field $\phi_i$.
If one has only Higgs singlets and doublets ($t_i = 0,\frac{1}{2}$),
then $\rho_0^{\rm tree} = 1$.
However, in the presence of larger representations with non-zero
vacuum expectation values
\beq \rho_0^{\rm tree} \simeq 1 + 2 \sum_i \left( t^2_i -
3 t_{3i}^2 + t_i \right) \frac{ |\langle \phi_i \rangle
|^2}{|\langle \phi_{\frac{1}{2}} \rangle |^2 }. \label{eq22a3}
\eeq

One can also have loop-induced contributions similar to that from $t$
and $b$, due to non-degenerate multiplets of fermions or bosons.  For new
doublets
\beq \rho_0^{\rm loop} = \frac{3G_f}{8 \sqrt{2} \pi^2} \sum_i
\frac{C_i}{3} F (m_{1i},m_{2i}),   \eeq
where $C_i = 3(1)$ for color triplets (singlets) and
\beq F(m_1, m_2) = m_1^2 + m^2_2 - \frac{4m_1^2 \; m^2_2}{m_1^2 -
m^2_2} \ln \frac{m_1}{m_2} \geq (m_1- m_2)^2 .  \label{f12def}  \eeq
Loop contributions to $\rho_0$ are generally positive\footnote{One can
have $\rho^{\rm loop} < 0$ for Majorana fermions \protect\cite{a34a} or
boson multiplets with vacuum expectation values \protect\cite{a34b}.}, and
if present would lead to lower values for the predicted $m_t$. $\rho_0^{\rm
tree}$ can be either positive or negative depending on the quantum numbers
of the Higgs field.  The $\rho_0$ parameter is extremely important because
one expects $\rho_0 \sim 1$ in most superstring theories~\cite{c1},
which generally
do not have higher-dimensional Higgs representations\footnote{The
only known exceptions are string
models in which the observed particles are composite, or models with
$k > 1$ worldsheet currents~\cite{v39}.},
while typically
$\rho_0 \neq 1$ from many sources in models involving compositeness.

In the presence of $\rho_0$ the standard model formulas for the observables
are modified.
As long as $\rho_0 - 1$ is
sufficiently small, one can simply incorporate the effects of $\rho_0
-1$ in the tree-level formulas and take $\rho_0 =1$ in the radiative
corrections~\cite{pom}.
At tree level,
\beq M_Z \ra \frac{1}{\sqrt{\rho_0}} M_Z^{SM}, \Gamma_Z \ra
\rho_0
\Gamma_Z^{SM}, {A}_{NC} \ra \rho_0 {A}^{SM}_{NC},\eeq
where ${A}_{NC}$ is a neutral current amplitude.
It has long been known that $\rho_0$ is close to 1. However, until recently it
has been difficult to separate $\rho_0$ from $m_t$, because most
observables only involve the combination $\rho_0 \hat{\rho}$.  The one
exception has been the $Z \ra b\bar{b}$ vertex.  However, assuming that CDF
has really observed the top quark directly one can use the known $m_t$ to
calculate $\hat{\rho}$ and therefore separate $\rho_0$.  In practice one
fits to $m_t$, $\rho_0$ and the other parameters simultaneously,
using the CDF value $m_t
= 174 \pm 16$~GeV as an additional constraint.  One obtains
\beq \begin{array}{ccc} \hat{s}^2_Z = 0.2316(3)(1) & \;\;\; & \rho_0 = 1.0012
\pm 0.0017 \pm 0.0017 \\ \alpha_s = 0.125(6)(1) & & m_t = 166 \pm 15 \pm 0 \
{\rm GeV}, \end{array} \label{eqrho} \eeq
where the second uncertanty is from $M_H$.  Even in the presence of the
classes of new physics parameterized by $\rho_0$ one still has robust
predictions for the weak angle and a good determination of $\alpha_s$.
Most remarkably, given the CDF constraint, $\rho_0$ is constrained to be
extremely close to unity, causing serious problems
for compositeness models.  The allowed region in $\rho_0$ vs $\hat{s}^2_Z$ are
shown in Figure \ref{figerica}.  This places limits $|\langle \phi_i
\rangle|/ |\langle \phi_{1/2} \rangle| < {\rm few} \%$ on non-doublet
vacuum expectation values, and places constraints $\frac{C}{3} F(m_1, m_2)
< O((100\; {\rm GeV})^2)$ on the splittings of additional fermion or boson
multiplets.

\begin{figure}
\postbb{60 230 535 680}{/home/pgl/fort/nc/graph/xrho/xxrhof.ps}{0.6}
\caption[]{90\% C.L. allowed regions
in $\rho_0$ vs $\hat{s}^2_Z$ for $M_H = 60$, 300,
and 1000 GeV.}
\label{figerica}
\end{figure}

One can also consider the possibility that there are both new sources
of $SU_2$ breaking and new contributions to the $Zb\bar{b}$ vertex.
A simultaneous fit to all data yields~\cite{a7c}
\beq \begin{array}{ccc} \hat{s}^2_Z = 0.2316(3)(2) & \;\;\; & \rho_0 = 1.0004
\pm 0.0018 \pm 0.0018 \\ \alpha_s = 0.111(9)(0) & & m_t = 174 \pm 16^{+1}_{-0}
\
{\rm GeV} \\ \delta^{\rm new}_{bb} = 0.022 \pm 0.011 \pm 0. &  \end{array} \eeq
Just as in the standard model (with $\rho_0 = 1$) the value of \als \
decreases significantly if one allow for a nonzero $Zb\bar{b}$ vertex
correction. The other parameters are changed moderately or not at all.

Even without the CDF constraint the special $m_t$ dependence
of $\Gamma(b\bar{b})$ can be used to obtain an upper limit on $m_t$
for arbitrary Higgs representations, i.e., arbitrary
$\rho_0$~\cite{rhonot,bb1}.
{}From the precision data and D0 lower bound ($m_t > 131$ GeV), one finds
$m_t < 183 (195)$ GeV at 90(95)\% C.L., essentially independent of $M_H$.

\vglue 0.6cm
\subsection{\elevenit Heavy Physics by Gauge Self Energies}
\vglue 0.4cm
\label{stusec}

A larger class of extensions of the standard model can be parameterized by
the $S$, $T$ and $U$ parameters \cite{a35}-\cite{a35e},
which describe that subset of
new physics which affect only the gauge boson self-energies\footnote{This
formalism assumes that the new physics is much
heavier than $M_Z$. The results can be generalized to
new physics scales comparable to $M_Z$ by introducing additional
parameters~\cite{burgess}.} but do not
directly affect tree-level amplitudes,
vertices, etc.  One introduces three parameters
\begin{eqnarray} S &=& S_{\rm new} + S_{m_t} + S_{M_H} \nonumber \\
                 T &=& T_{\rm new} + T_{m_t} + T_{M_H}  \\
   U &=& U_{\rm new} + U_{m_t}. \nonumber \end{eqnarray}
The new physics contributions are defined by
\begin{eqnarray} \alpha T_{\rm new}^\rmm{loop}
&\equiv & \frac{\Pi^{\rm new}_{WW}(0)}{M_W^2} -
\frac{\Pi^{\rm new}_{ZZ}(0)}{M_Z^2}
\nonumber \\
\frac{\alpha}{4 \shz \hat{c}^2_Z} S_{\rm new} & \equiv &
\frac{\Pi^{\rm new}_{ZZ}(M^2_Z) - \Pi^{\rm new}_{ZZ}(0)}{M_Z^2}  \\
\frac{\alpha}{4 \shz} (S+U)_{\rm new} & \equiv &
\frac{\Pi^{\rm new}_{WW}(M^2_W) - \Pi^{\rm new}_{WW}(0)}{M_W^2}  \\
   \nonumber \end{eqnarray}
where $\Pi^{\rm new}_{WW}$ and $\Pi^{\rm new}_{ZZ}$ are the contributions
of new physics to the $W$ and $Z$ self-energies.
$T$ is associated with the difference between the $W$ and
$Z$ self-energies at $Q^2= 0$ and describes
the breaking of the $SU_{2V}$ vector generators.
 $T$ is equivalent to the
$\rho_0$ parameter and is induced by mass splitting in multiplets of
fermions or bosons.
$S$ ($S+U$) are associated with the differences between the
$Z$ ($W$) propagators at $Q^2 = 0$ and $M^2_Z \ (M^2_W)$, and
describe the breaking of the $SU_{2A}$ axial generators. $S$ is
generated, for example, by degenerate heavy chiral families of fermions.
$U$ is zero in most extensions of the standard model.
$S$, $T$, and
$U$ are induced by loop corrections and have a factor of $\alpha$ extracted,
so they are expected to be $O(1)$ if there is new physics.
They are related to equivalent parameters
defined in~\cite{a35a} by
\begin{eqnarray} S &=& h_{AZ} = S_Z = 4 \shz {\epsilon}_3/\alpha
\nonumber \\
                 T &=& h_{V} =   {\epsilon}_1/\alpha \\
   U &=& h_{AW} - h_{AZ} = S_W - S_Z = -4 \shz {\epsilon}_2/\alpha.
   \nonumber \end{eqnarray}

$S$,
$T$ and $U$ were introduced to describe the contributions of new physics.
However, they can also parametrize the effects of very heavy $m_t$ and
$M_H$ (compared to $M_Z$). Expressions for $S_{m_t}$ and $S_{M_H}$, which
are respectively the $m_t$ and $M_H$ contributions to $S$, and
similarly for $T$ and $U$, may be found in ~\cite{a35a,a35d}.
Until recently it was difficult to separate the
$m_t$ and new physics contributions. Therefore, most analyses fixed
$m_t$ and $M_H$ at arbitrary reference values (e.g., $M_Z$, or
the result of the best fit in the standard model), and fit to the
total $S$, $T$, and $U$. The results could then be compared to the
standard model expectations for other values of $m_t$ and $M_H$.
Now, however, with the CDF value of
$m_t$ it is possible to
directly extract the new physics contributions.
That is, one can determine $S_{\rm new}$, $T_{\rm new}$,
and $U_{\rm new}$ in a simultaneous fit with \shz, $m_t$, \als,
and (optionally) $\delta^{\rm new}_{bb}$, with the $M_H$ dependence
included in the uncertainties. In practice, one can use the full $m_t$
and $M_H$ dependence of all observables, and not just their
contributions to $S$, $T$, and $U$, which are approximations valid for
masses much larger than $M_Z$.

A new multiplet of degenerate chiral fermions will contribute to $S_{\rm
new}$ by
\beq S_{\rm new} |_{\rm degenerate} = \sum_i C_i |t_{3L} (i) - t_{3R} (i)
|^2/3\pi \geq 0, \eeq
where $C_i$ is the number of colors and $t_{3LR}$ are the $t_3$ quantum
numbers.  A fourth family of degenerate fermions would yield
$\frac{2}{3\pi} \sim 0.21$, while QCD-like technicolor models, which
typically have many particles, can give larger contributions.  For example,
$S_{\rm new} \sim 0.45$ from an isodoublet of fermions
with four technicolors, and an
entire technigeneration would yield $1.62$ \cite{a35}.  Non-QCD-like theories
such as those involving walking could yield smaller or even
negative contributions \cite{a37}.  Nondegenerate scalars or fermions can
contribute to $S_{\rm new}$ with either sign \cite{a38}.

The $T$ parameter is analogous to $\rho_0^{\rm loop}$.  For a
non-degenerate family
\beq T_{\rm new}^\rmm{loop}
\sim \frac{\rho_0^{\rm loop}}{\alpha} \sim 0.42
\frac{\Delta m^2}{(100 \;GeV)^2}, \eeq
where
\beq \Delta m^2 = \sum_i \frac{C_i}{3} F \left( m_{1i}, m_{2i} \right) \geq
\sum_i \frac{C_i}{3 } \left( m_{1i} - m_{2i} \right)^2 \eeq
and $F(m_1,m_2)$ is defined in (\ref{f12def}).
Usually $T_{\rm new}^\rmm{loop} > 0$,
although there may be exceptions for theories
with Majorana fermions~\cite{a34a} or additional
Higgs doublets~\cite{a34b}.  In practice,
higher-dimensional Higgs multiplets could mimic $T_{\rm new}$ with either
sign (see equation (\ref{eqerica})), and cannot be separated from loop
effects unless they are seen directly or have other effects.
That is, $T_\rmm{new}^\rmm{loop}$ and the contribution $\rho_0^{\rm tree}$
enter
observables in the universal combination $\rho_0^{\rm tree}/(1 -
\alpha T_\rmm{new}^\rmm{loop})$. Therefore, although $T$ was originally
defined to include loop contributions only, I will extend the definition
to include tree-level effects,
\beq T_\rmm{new} \equiv T_\rmm{new}^\rmm{loop} + T_\rmm{new}^\rmm{tree}, \eeq
where $\rho_0^{\rm tree} = 1 + \alpha T_\rmm{new}^\rmm{tree}$.
Then,
\beq \rho_0 = 1 + \alpha T_\rmm{new} \label{rhoandt} \eeq
are equivalent parameters describing  new sources of both tree and loop level
$SU_{2V}$ breaking.

Usually $U_{\rm new}$ is small, although there are counterexamples,
such as anomalous triple-gauge vertices~\cite{a35e}.
Supersymmetric extensions of the standard model usually
give negligible contributions to $S$, $T$, and $U$~\cite{susyrad}.

The standard model expressions for observables are replaced by
\bqa  M_Z^2 &= & \frac{1}{\rho_0} \left( M_{Z}^{SM} \right)^2
\frac{1}{1 - G_F \left( M_{Z}^{SM} \right)^2 S_\rmm{new}/2 \sqrt{2} \pi}
\nonumber \\
M_W^2 &= & \left( M_{W}^{SM} \right)^2
\frac{1}{1 - G_F \left( M_{W}^{SM} \right)^2
(S + U)_\rmm{new}/2 \sqrt{2} \pi},  \eqa
where $M_{W,Z}^{SM}$ are the standard model expressions
(\ref{eq2},\ref{eq3}) in
terms of \shz, $m_t$, $M_H$, and \als, and $\rho_0$ is related
to $T_\rmm{new}$ by (\ref{rhoandt}). Furthermore,
\bqa \Gamma_Z &= & \rho_0  M_Z^3 \beta_Z
\nonumber \\
\Gamma_W& = &M_W^3 \beta_W   \nonumber \\
A = &= & \rho_0  A^{SM},  \eqa
where $\beta_{Z,W}$ is the standard model expression for the reduced
width $\Gamma_{Z,W}^{SM}/\left( M_{Z,W}^{SM} \right)^3$,
$M_{Z,W}$ is the physical mass, and
$A$ ($A^{SM}$) is a neutral current amplitude (in the standard model).

There is enough data to simultaneously determine the new physics
contributions to $S$, $T$, and $U$, the standard model parameters, and also
$\delta^{\rm new}_{bb} =
\frac{\Gamma (b\bar{b})}{\Gamma^{\rm SM} (b\bar{b})} -1
$. For example, $S_{\rm new}$, $T_{\rm new}$, $U_{\rm new}$, $\delta^{\rm
new}_{bb}$, $\hat{s}^2_{Z}$, $\alpha_s(M_Z)$ and $m_t$ are constrained by
$M_Z$, $\Gamma$, $M_W$, $R_b$, asymmetries, $R$, and $m_t$ (CDF),
respectively.  One obtains~\cite{a7c}
\begin{eqnarray} S_{\rm new} = -0.21 \pm 0.24^{-0.08}_{+0.17} &
\; \; \;\;\; & \hat{s}^2_Z = 0.2313 (4) (1) \nonumber \\
T_{\rm new} = -0.09 \pm 0.32^{+0.16}_{-0.11} &
\; \; \;\;\; & \alpha_s(M_Z) = 0.112 (9)(0) \nonumber \\
U_{\rm new} = -0.53 \pm 0.61 & \; & m_t = 175 \pm 16 \pm 0
\; {\rm GeV} \\
\delta^{\rm new}_{bb} = 0.022 \pm 0.011 \pm 0, & \; & \; \nonumber
\end{eqnarray}
where the second error is from $M_H$.
The $T_{\rm new}$ value corresponds to
$\rho_0 = 0.9993 \pm 0.0023 ^{+ 0.0012}_{- 0.0008}$, which differs
from the value in (\ref{eqrho}) because of the presence of $S_{\rm new}$,
$U_{\rm new}$, and $\delta^{\rm new}_{bb}$.
The data is consistent with the
standard model: $S_{\rm new}$ and $T_{\rm new}$ are close to zero with
small errors, and the tendency to find $S < 0$ that existed
in earlier data is no longer present (although $A^0_{LR}$ by itself
favors $S < 0$).
The constraints on $S_{\rm new}$ are a problem for those
classes of new physics such as technicolor which tend to predict $S_{\rm new}
$ large and positive, and $S_{\rm new}$ allows, at most, one additional
family of ordinary fermions at 90\% CL.  (Of course the invisible $Z$ width
precludes any new families unless the additional neutrinos are heavier than
$M_Z/2$.)  The allowed regions in $S_{\rm new}$ vs $T_{\rm new}$
are shown in Figure~\ref{fig5}.
The seven parameter fit still favors a non-zero $Z\ra b\bar{b}$ vertex
correction $\delta^{\rm new}_{bb}$, almost identical to the value
obtained without $S$, $T$, and $U$. The low value of the extracted
\als \ compared to the standard model value $(0.127 (5)(2))$ is entirely
due to $\delta^{\rm new}_{bb}$.
The value of $\hat{s}^2_Z$ is slightly lower than the standard model value
$(0.2317 (3)(3))$.
One can repeat the fits without the $Zb\bar{b}$ correction
$\delta^{\rm new}_{bb}$. One finds almost identical values for
$S_\rmm{new}$, $T_\rmm{new}$, and $U_\rmm{new}$. Now, however,
\als = 0.125 (6)(0), close to the standard model value.

\begin{figure}[tbh]
\postbb{60 230 520 670}{/home/pgl/fort/nc/graph/st/xxst.ps}{0.7}
\caption[]{Constraints on $S_{\rm new}$ and $T_{\rm new}$ from various
observables and from the global fit to all data. The fit to
$M_W$ and $M_Z$ assumes $U_\rmm{new} = 0$, while $U_\rmm{new}$
is free in the other fits.}
\label{fig5}
\end{figure}

There is no simple parametrization which utilizes all
experimental information and which describes every
type of new physics~\cite{a35d}.
The $S$, $T$, and $U$ formalism parametrizes important classes
of new physics associated with gauge boson self-energies and
non-standard Higgs representations. It utilizes all experimental
information, but does not apply to such effects as non-universal
vertex corrections.
An alternative formalism~\cite{a35c} is based
on the shifts induced in $M_W/M_Z$, $\Gamma(\ell \bar{\ell})$,
$A_{FB}^{0\ell}$, and $R_b$. It applies to all types of new physics,
but cannot make use of other observables unless extra assumptions
are made. A more general possibility involves
deviation vectors (M. Luo, this volume, and \cite{LLM}),
as shown in Figure \ref{devvec}.
Each type of new physics defines a deviation vector, the components
of which are the deviations of each observable from its Standard
Model prediction, normalized to the uncertainty. The length
(direction) of the vector represents the strength (type) of
new physics.
The latter would be expecially
convenient for diagnosing the origin of new physics if
significant deviations were observed. One can also describe
new physics by effective Lagrangian techniques~\cite{effective},
which are especially useful for enforcing  the correlated effects of
different operators that are related by gauge invariance or
other symmetries.
Many types of new physics,
such as heavy $Z^\prime$ bosons or mixing with exotic fermions,
are most conveniently described by special parametrizations.
\begin{figure}[tbh]
\postbb{60 250 530 680}{/home/pgl/fort/nc/graph/dev/xxdev.ps}{0.6}
\caption[]{Deviation vectors for various observables. Each bar
represents $(O_i - O^{SM}_i)/\Delta O_i$, where $O_i$ is the
experimental value of the $i^{th}$
observable, $O^{SM}_i$ is the standard model
prediction for $O_i$ in terms of $M_Z$ and the global allowed
ranges of $m_t$, \als, and $M_H$, and $\Delta O_i$ is the
total uncertainty, obtained by adding the experimental
and theoretical uncertainties in quadrature.}
\label{devvec}
\end{figure}

\vglue 0.6cm
\subsection{\elevenit Additional $Z^\prime$ Bosons}
\vglue 0.4cm

\begin{figure}
\small \def\baselinestretch{1} \normalsize
\postbb{60 210 530 670}{/home/pgl/fort/nc/graph/z/xxzchi.ps}{0.6}
\caption[]{
90\% CL $(\Delta \chi^2 = 4.6$) allowed region in $M_2$ and $\theta$
for the $SO_{10}$ boson $Z_{\chi}$ using indirect (WNC, $M_{W,Z}$,
$Z$-pole)
data for the cases $\rho_0$-free and $\rho_0 = 1$.
\als, $m_t$, and \shz \ are free parameters, $\lambda_g = 1$ is assumed,
and the CDF constraint $m_t = 174 \pm 16$ GeV is included.
Also shown are the additional constraint in the minimal Higgs case, best
fit point, and the band of 95\% CL lower limits on $M_2$ from the direct CDF
search.  Updated from \protect\cite{v11}.}
\label{chi}
\end{figure}

Many extensions of the standard model predict the existence of
additional gauge bosons, which could
be light enough to be observable \cite{n2a}-\cite{n2d}.
For example, there is an extra $Z'$
in $SU_{2L} \x SU_{2R} \x U_1$ models, which couples to the
current~\cite{n2a}
\beq J_{LR} = \sqrt{\frac{3}{5}} \left( \alpha J_{3R} -
\frac{1}{2\alpha} J_{B-L} \right), \label{eq328.3} \eeq
where $T_{3R}$ is the third component of $SU_{2R}$, $B-L$ is the $U_1$
current
\beq B-L = 2 (Q - T_{3L} - T_{3R}), \label{eq401.1} \eeq
and
\beq \alpha = \left[ \left( \frac{g_R}{g_L} \right)^2 \left( \frac{1 -
\sin^2 \theta_W}{\sin^2\theta_W} \right) -1 \right]^{1/2},
\label{eq327.5} \eeq
which is $\simeq 2.53$ for left-right symmetry $(g_R = g_L)$.
Similarly, many grand unified theories lead to extra $U_1$'s when they
break, such as~\cite{gutpat,Z}
\beqa SO_{10} & \ra & SU_5 \x U_{1\chi} \nonumber \\
E_6 & \ra & SO_{10} \x U_{1 \psi} \nonumber \\
E_6 & \ra & SU_3 \x SU_2 \x U_1 \x U_{1 \eta} \label{eq328.6} \eeqa
The corresponding charges are shown in Table~\ref{Za}.
\begin{table}                    \centering
\small \def\baselinestretch{1} \normalsize
 \begin{tabular}{|ccccc|}
        \hline
  $ SO_{10} $  & $ SU_{5} $ & $ 2 \sqrt{10} Q_{\chi} $ &
  $ \sqrt{24} Q_{\psi} $ & $ 2 \sqrt{15} Q_{\eta} $ \\
 \hline
16  & 10 $(u,d,\bar{u},e^{+})_{L}$  & $-1$ & $1$ & $-2$ \\
    & $ 5^{*} (\bar{d},\nu, e^{-})_{L}$  & $3$ & $1$ & $1$ \\
    & $ 1 \bar{N}_{L}$  & $-5$ & $1$ & $-5$ \\
10  & 5 $(D,\bar{E^{0}},E^{+})_{L}$  & $2$ & $-2$ & $4$ \\
    & $ 5^{*} (\bar{D},E^{0},E^{-})_{L}$  & $-2$ & $-2$ & $1$ \\
1   & $ 1  S^{0}_{L}$  & $0$  & $4$  & $-5$ \\
        \hline
\end{tabular}
\caption[]{Couplings of the $Z^{0}_{\chi}$, $Z^{0}_{\psi}$, and
 $Z^{0}_{\eta}$ to a 27-plet of $E_{6}$. The $SO_{10}$ and $SU_{5}$
 representations are also indicated. The couplings are shown for
 the left-handed ($L$) particles and antiparticles. The couplings of
 the right-handed particles are minus those of the corresponding
 $L$ antiparticles. The $D$ is an exotic $SU_{2}$-singlet quark
 with charge $-1/3$. $(E^{0}, E^{-})_{L,R}$ is an exotic lepton
 doublet with vector $SU_{2}$ couplings. $N$ and $S$ are new Weyl
 neutrinos which may have large Majorana masses.}
\label{Za}
\end{table}
The $Z_\chi$ couplings are a special case of the $Z_{LR}$,
corresponding to $g_R = g_L$ and $\sinn = 3/8$ (\ie $\alpha
= \sqrt{2/3} = 0.82$, while
\beq J_\eta = \sqrt{\frac{3}{8}}  J_\chi - \sqrt{\frac{5}{8}} J_\psi
\label{eq328.7} \eeq
emerges in a particular $E_6$-breaking that is motivated by some
superstring-inspired models.  The gauge coupling $g_2$ of the extra
$Z$ is given by
\beq \left( \frac{g_2}{g_1} \right)^2 = \frac{5}{3} \sin^2 \theta_W
\lambda_g, \label{eq329.1} \eeq
where $g_1 = g/\cos\theta_W$ and $\lambda_g \leq 1$ depends on the
symmetry-breaking pattern~\cite{gutpat} ($\lambda_g = 1$ by
construction for the $Z_{LR})$.

\begin{figure}
\small \def\baselinestretch{1} \normalsize
\postbb{60 210 530 670}{/home/pgl/fort/nc/graph/z/xxzpsi.ps}{0.6}
\caption[]{Same as \protect\ref{chi}, only for $Z_\psi$. In the
minimal-Higgs case, $\sigma = 0$, 1, 5, and $\infty$.}
\label{psi}
\end{figure}

The above models all involve family-universal fermion couplings,
which ensures the absence of flavor changing
neutral currents. The possibility of a $Z'$ which couples only
to the third family is discussed in~\cite{nuzp}.

In the presence of one extra $Z$, there are two observed mass
eigenstate bosons $Z_1$ and $Z_2$,
with masses $M_1$, $M_2$, which are
related to the weak eigenstates $Z^0_{1,2}$
by a mixing angle $\theta$:
\beqa Z_1 &=& Z_1^0 \cos \theta + Z_2^0 \sin \theta \nonumber \\
 Z_2 &=&- Z_1^0 \sin \theta + Z_2^0 \cos \theta .\label{eq329.2} \eeqa
I assume that $Z_1$ is the observed $Z$.  The $Z^0_1$ is the $SU_2 \x
U_1$ model boson, and the $Z^0_2$ has the couplings of the new $U'_1$
group.  The presence of the extra $Z$ modifies the neutral current and
$Z$-pole results in three ways: (a) the $Z_1$ mass $M_1$ is reduced
compared to the standard model prediction due to mixing, (b) the $Z_1$
couplings are changed by mixing (this can be especially important for
$g_{Ve}$, which is small in the standard model), and (c) $Z_2$ exchange
contributes to neutral current processes.  All of these must be
included in global fits to the data\footnote{A further extension, in
which one allows for the simultaneous effects of an extra $Z$
and mixing between ordinary and exotic fermions, is discussed
in ~\cite{zexotic}.}.

Indirect limits on the mass $M_2$ and mixing angle $\theta$ can be
obtained from the $Z$, $W$, and neutral current data.  Because of
large number of possible couplings to fermions it is easiest to
consider specific models, such as the $Z_\chi$, $Z_\psi$, $Z_\eta$,
and $Z_{LR}$.  One can consider three versions of each model, depending
on how much is assumed concerning the Higgs sector of the theory.

In each case, one has~\cite{2zmix}
\beq \tan^2\theta = \frac{M_0^2 - M_1^2}{M_2^2 - M_0^2},
\label{mix} \eeq
where $M_{1,2}$ are the physical $Z_{1,2}$ masses, and $M_0$ is the
$Z^0_1$ mass before mixing, {\it i.e.,}
\beq M_0 = \frac{M_W}{\sqrt{\rho_0 \hat{\rho}} \hat{c}_Z} =
\frac{M_W}{\sqrt{\rho_0} c_W}, \label{m0} \eeq
where $\rho_0$ is given by (\ref{eqerica}), just as in the standard
model. Eqn.~(\ref{m0}) is written at tree-level. The normal $SU_2\x U_1$
radiative corrections must of course be applied to all formulae.
Since, we are searching for small tree-level perturbations, it is a
good first approximation\footnote{Some care must be applied when using
the on-shell definition~\protect\cite{degsir}, and there is a small
correction due to the $Z_2$ contribution to muon
decay~\cite{marsir}.} to ignore
contributions of the extra $Z$ to the radiative corrections.

 In the unconstrained-Higgs $(\rho_0$ free) version, no assumption is
made concerning the quantum numbers of the Higgs fields which break
the $SU_2 \x U_1 \x U'_1$ symmetry.  This is analogous to the $\rho_0
\neq 1$ extension of the standard model (due to Higgs triplets, etc.).
In this case, $\rho_0$ and hence $M_0$ are unconstrained, and \shz,
$M_1$, $M_2$, and $\theta$ are all free
parameters (as are $m_t$ and \als).  There are so many experimental
constraints, however, that these can all be simultaneously
determined or constrained. The   90 and  95\% CL lower limits on $M_2$
for
the $\chi$, $\psi$, $\eta$, and $LR$ models and for a heavy $Z'$ with the
same couplings as the ordinary $Z$ (such couplings are not expected in
gauge theories, but are useful for comparison) are shown in
Table~\ref{zmasslim}, and the allowed regions in $M_2$ and $\theta$ in
Figures~\ref{chi}--\ref{lr}.  The limits on the mixing angle $\theta$
are shown for each model in Table~\ref{ztheta}.
In all fits, the CDF constraint $m_t = 174 \pm 16$ GeV is included.
The values of \shz, \als, $m_t$, and $\rho_0 - 1$ are almost identical to the
$SU_2 \times U_1$  case.
\begin{table}   \centering
\begin{tabular}{|r|rrrrr|}
\hline
& $Z_{\chi}$ & $Z_{\psi}$ & $Z_{\eta}$ & $Z_{LR}$ & $Z^{\prime}$ \\
\hline
$\rho_{0}=$ free & 353(378) & 167(181) & 216(237) & 389(420) &
951(1050)\\
$\rho_{0}=1$ & 334(352) & 168(182) & 217(237) & 391(422) &
958(1050)\\
\hline
$\sigma = 0$ & 919(1020) & 954(1040) & 407(436) & 1360(1470) & -- \\
$\sigma = 1$ & -- & 167(181) & 614(673) & -- & -- \\
$\sigma = 5$ & -- & 791(851) & 930(1010) & -- & -- \\
$\sigma = \infty$ & -- & 1020(1100) & 1090(1190) & -- & --
\\
\hline
direct      & 360-425 & 280-415 & 290-440 & 350-445 & 505 \\
\hline
\end{tabular}

\caption[]{Lower limits on the mass $M_2$ of an extra $Z$ boson for the
$Z_{\chi}, Z_{\psi}, Z_{\eta}, Z_{LR}$, and $Z^{\prime}$ models. The
indirect limits are for the $\rho_0=$free, $\rho_0=1$, and for minimal
Higgs models with $\sigma = 0, 1,5,\infty$. (For the $Z_{\chi}$ and $Z_{LR}$
the limits are independent of $\sigma$.) Both the 95\% CL and 90\%
(in parentheses) CL limits are given.  In all cases,
$\shz$, $m_t$, and \als \ are free parameters, and the CDF constraint
$m_t = 175 \pm 16$ GeV is included.
The direct limits are 95\% CL based on the
CDF search for $\bar{p}p \RA Z' \RA e^+ e^-$ \cite{z1}.
The range of values for each model is for various possibilities for
$B (Z_2 \rightarrow e^+ e^-$), ranging from the
assumption that the only open decay channels are the known fermions
(strongest limits), to the case that decays can also occur into exotic
fermion and superpartner channels.
Updated from \protect\cite{v11}.}
\label{zmasslim}
\end{table}

\begin{figure}
\small \def\baselinestretch{1} \normalsize
\postbb{60 210 530 670}{/home/pgl/fort/nc/graph/z/xxzeta.ps}{0.6}
\caption[]{Same as \protect\ref{psi}, only for $Z_\eta$.}
\label{eta}
\end{figure}

\begin{figure}
\small \def\baselinestretch{1} \normalsize
\postbb{60 210 530 670}{/home/pgl/fort/nc/graph/z/xxzlr.ps}{0.6}
\caption[]{Same as \protect\ref{chi}, only for $Z_{LR}$, assuming
$g_R = g_L$.}
\label{lr}
\end{figure}

\begin{table}   \centering
\begin{tabular}{|r|rrrrr|}
\hline
& $Z_{\chi}$ & $Z_{\psi}$ & $Z_{\eta}$ & $Z_{LR}$ & $Z^{\prime}$ \\
\hline
$\rho_0=$ free & --0.0025 (26) & --0.0005(25) & $-$0.0024(35)
& 0.0003(16) & $-$0.0007(12)\\
$\theta_{\rm{min}}$      & $-$0.0067   & $-$0.0046  & $-$0.0080
& $-$0.0024  & $-$0.0027\\
$\theta_{\rm{max}}$      & +0.0017     & $+$0.0036   & +0.0034
& $+$0.0029  & +0.0013\\
\hline
$\rho_0=1$       &--0.0026(19)   & 0.0000$^{+20}_{-32}$ & $-$0.0023(34)
& 0.0003(18) & $-$0.0006(12)\\
$\theta_{\rm{min}}$      & $-$0.0054   & $-$0.0046  & $-$0.0079
& $-$0.0025  & $-$0.0027\\
$\theta_{\rm{max}}$      & +0.0013     & +0.0036     & +0.0033
& +0.0029    & +0.0010\\
\hline
\end{tabular}
\caption[]{ Best fit values and 95\% CL upper ($\theta_{\rm{max}}$) and
lower ($\theta_{\rm{min}})$ limits on the mixing angle $\theta$ for the
$\rho_0=$ free and $\rho_0 =1$ models. The numbers in parentheses are
the uncertainties in the best fit values. Updated from \protect\cite{v11}.}
\label{ztheta}
\end{table}

In the constrained-Higgs ($\rho_0 = 1$) case one assumes that
$SU_{2L}$ breaking is due to Higgs doublets, but no assumption is made
concerning their $U'_1$ charges.  In this case, $\rho_0 =1$, so $M_0$
is known.  The free parameters are \shz, $M_2$,
$\theta$ (and $m_t$ and \als).  $M_1$ is not independent, but is given by
(\ref{mix}).  From Tables~\ref{zmasslim} and \ref{ztheta} and
the Figures one sees that
the $M_2$ and $\theta$ limits are comparable to the unconstrained case.

Finally, there are the minimal-Higgs fits, in which one assumes not
only that the relevant Higgs fields are $SU_2$ doublets and singlets,
but also specific values for their $U'_1$ charges.  Since the same
Higgs fields implement both $SU_2$ breaking and $Z-Z'$ mixing, such
models lead to a relation
\beq \theta = C \frac{M^2_1}{M_2^2}, \label{eq329.4} \eeq
where $C$ depends on the Higgs quantum numbers.  In some cases, such as
the $Z_\chi$ model and some $Z_{LR}$ models\footnote{Those for
which the $SU_{2R}$ breaking is due to Higgs
triplets~\cite{lrmodels}.}, $C$ is
a definite prediction, and in others, such as $Z_\psi$ and
$Z_\eta$, it varies over a
finite range
depending on the relative expectation values $\sigma$ of two Higgs
doublets.
The predictions for $C$ in terms of $\sigma = |\bar{v}|/|v|$
are given in
\cite{v11}. It is expected that
$\sigma > 1$, since $\bar{v}$ and $v$ give rise to $m_t$ and $m_b$,
respectively.
{}From Table~\ref{zmasslim} and the Figures it is apparent that one often
gets much more stringent limits than in the ordinary $\rho_0 =1$ case.

There are also direct limits on $Z$'s from CDF at the Tevatron, from
searches for $\bar{p}p \ra Z' \ra e^+e^-$.  The upper limit
depends on the $Z'$ mass, but for the 500 GeV range
is around~\cite{z1}
\beq \sigma_{\bar{p}p\ra Z'} \;\; B_{Z' \ra e^+e^-} \; < 0.33 \;pb
\label{eq329.5} \eeq
at 95\% CL at $\sqrt{s} = 1.8$~TeV.  This would correspond to a limit
of $M > 505$~GeV for a $Z'$ with the same couplings as the ordinary
$Z$, but a weaker limit on more realistic $Z$'s.  It is safe to ignore $Z-Z'$
mixing in the
analysis, since indirect limits indicate that $\theta$
is small.  The resulting limits on the $\chi$, $\psi$, $LR$, and
$\eta$ are shown in Table~\ref{zmasslim} and in the figures.
A range of limits is displayed, corresponding to different
assumptions concerning the $Z' \RA e^+ e^-$ branching ratio,
\ie whether the only open channels are the known fermions (strongest
limits), or whether decays into exotic fermions and superpartners
are allowed (weakest limits).
The direct limits are currently slightly stronger than the
indirect limits for the unconstrained and constrained Higgs
cases, but weaker than most of the limits for the minimal-Higgs
models.

Altogether, the current limits on the masses of extra $Z$'s are rather
weak (typically $\sim$ 400 GeV) except when specific Higgs representations are
assumed, and depend strongly on the $U'_1$ charges.  However, there are
rather strong
limits on the mixing angles, mainly from the $Z$-pole data.
Further implications are discussed in \cite{v11,v11a},
future prospects for indirect searches are considered
in~\cite{LLM}, and the direct discovery and diagnostic
potential of future hadron and $e^+ e^-$ colliders
in~\cite{v11b}.

\vglue 0.6cm
\subsection{\elevenit Exotic Fermions}
\vglue 0.4cm
\label{exotic}
The known fermions are all ordinary \cite{v35}, \ie the left-handed
particles transform as weak doublets, and the right-handed
particles as singlets:
\beq \doub{\nu_e}{e^-}{L},\ \ \doub{u}{d}{L}, \ \ e^-_R, \ \ u_R, \ \
  d_R \eeq
It is of course possible that there are additional
(sequential) families of ordinary fermions. However, the
LEP constraint $N_\nu = 2.988 \pm 0.023$~\cite{a1} from the invisible
$Z$ width excludes sequential families unless the additional
neutrinos are far heavier ($m_\nu \simgr $ 40 GeV) than those of the
known families. Another constraint comes from the $S_\rmm{new}$ parameter,
\ie the shift in $M_Z$ induced by vacuum polarization diagrams
involving the new families (see section \ref{stusec}). From the
current value $S_\rmm{new} = -0.21 \pm 0.24^{-0.08}_{+0.17}$
and the contribution
\beq \Delta S = \frac{2}{3\pi} [ (N_L -3) + N_R] \eeq
of $N_L$ ordinary families and $N_R$ mirror (right-handed doublet,
left-handed singlet) families, we obtain the 90\% CL limit
\beq N_L - 3 + N_R < 1.6, \eeq
\ie it is unlikely that there are more than 1 or 2 additional families,
even if the associated neutrinos are very heavy.
The constraints on exotic fermions from both charged and
neutral current processes are described in detail by London and
by Herczeg in this volume, and in~\cite{v35}.

\vglue 0.6cm
\subsection{\elevenit Four-Fermi Operators and Leptoquarks}
\vglue 0.4cm

\label{fourferm}
Many types of new physics, such as compositeness, dynamical
symmetry breaking, leptoquarks, and supersymmetry lead to
new effective four-fermi operators (in
supersymmetry they are expected to be small for the interesting models).
These generally lead to
flavor changing neutral current effects. Even if these are
somehow evaded, there are often flavor-conserving
effects which contribute to charged and neutral current
processes, universality violation, etc.~\cite{LLM,r10,a30}.

The $Z$-pole observables, although most precise, are only
sensitive to the properties of the $Z$ and are essentially
blind to such operators\footnote{On the $Z$-pole, new operators
are out of phase and do not interfere with the $Z$ amplitude. Off
peak there may be interference, but the $Z$ amplitude is
suppressed.}.
However, neutral current processes are often quite sensitive. In
particular, atomic parity violation (APV) is sensitive to those new
operators which shift the values of $C_{1u}$ and $C_{1d}$ in
(\ref{eqch124}) in a direction orthogonal to the standard model
band in Figure~\ref{fig49}.

As an example, consider the effective operator
\beq  L_\rmm{new} = \pm \frac{4 \pi}{\Lambda^2_\pm}
  \bar{e}_L \gamma_\mu e_L  \bar{q}_L \gamma^\mu q_L, \eeq
  which shifts $C_{ij}$ by
  \beqa \Delta C_{1u} & = & \Delta C_{1d} = \Delta C_{2u} = \Delta C_{2d}
      = \frac{\mp \sqrt{2} \pi}{G_F \Lambda^2_\pm} \nonumber \\
       & = & -0.0029 \pm 0.0023, \eeqa
       where the second line is the constraint from current
       data~\cite{LLM,r10}.
This implies
\beqa  \Lambda_+ & < & 7.6 \ \rmm{TeV}  \nonumber \\
       \Lambda_- & < & 21 \ \rmm{TeV},  \eeqa
at 95\% C.L. This is already a stronger limit than will be
be attainable at HERA, although HERA would also be sensitive to
parity-conserving operators. Future APV experiments (see Masterson and
Wieman, this volume) will be even more sensitive.

As another example, consider a charge $-\frac{1}{3}$ scalar
leptoquark $S$, such as in predicted by some grand unified
theories~\cite{LLM}, with an interaction
\beq  -L = \left[ \eta_L (\bar{u}^c_R e_L - \bar{d}^c_R \nu_L) +
   \eta_R \bar{u}^c_L e_R \right] S + H.C.  \eeq
This implies
\beqa \Delta C_{1u} & = & \Delta C_{2u} =
  \frac{\mp \sqrt{2} \eta^2_{L,R}}{8 G_F M^2_S} \nonumber \\
  & = & -0.0064 \pm 0.0047,  \eeqa
  where  the current experimental constraint
  in the second line implies
\beq  M_S > \left\{ \begin{array}{c}
315 \ \ \rmm{GeV} | \eta_L |/\sqrt{4 \pi \alpha} \\
1040 \ \rmm{GeV} | \eta_R |/\sqrt{4 \pi \alpha}.
\end{array} \right. \eeq
(Universality constraints imply that either $\eta_L$ or $\eta_R$
should be negligibly small for relevant $M_S$.) These limits
equal or exceed what will be attainable at HERA for
electromagnetic-strength leptoquark couplings, although
HERA will be sensitive to light ($M_S < O(300 \ \rmm{GeV})$)
leptoquarks even for very small couplings. The APV limits on $M_S$
should be improved by a factor of 4 or so.
There are also significant constraints on leptoquarks
from $Z\ell\bar{\ell}$ vertex corrections~\cite{leptovertex}.
Leptoquarks are
discussed in more detail in this volume by Herczeg and
by Deutsch and Quin.

\vglue 0.6cm
\section{\elevenbf Conclusions}
\vglue 0.4cm

\begin{itemize}

\item The precision data have confirmed the standard electroweak model.
However, there are possible hints of discrepancies at the 2 -- 3 $\sigma$
level in $\Gamma (b\bar{b})/\Gamma(\rm had)$ and $A^0_{LR}$.

\item The data not only probes the tree-level structure, but the
electroweak loops have been observed at the $2\sigma$ level.  These consist
of much larger fermionic pieces involving the top quark and QED, which only
partially cancel the bosonic loops.  The bosonic loops, which probe
non-abelian vertices and gauge-Higgs vertices, are definitely needed to
describe the data.

\item The global fit to the data within the standard model yields
\beq \begin{array}{ccc} \overline{MS}: \hat{s}^2_Z = 0.2317 (3)(2) &
\;\;\;\; & m_t = 175 \pm 11\,^{+17}_{-19}\\ {\rm on-shell:} \; s^2_W
\equiv 1 - \frac{M_W^2}{M_Z^2}  = 0.2243 (12) & & \alpha_s (M_Z)
= 0.127 (5)(2) \\
{\rm effective:} \; \bar{s}^2_\ell = 0.2320 (3) (2) & & \end{array} \eeq
where the second uncertainty is from $M_H$.  The prediction for $m_t$ is in
remarkable agreement with the value $m_t = 1 74 \pm 16$ suggested by the
CDF events.  The data has also allowed, for the first time, a clean and
precise extraction of $\alpha_s$ from the lineshape.  This is in excellent
agreement with the value $\alpha_s (M_Z) = 0.123 \pm 0.005$ from event
shapes.  Both are larger than many of the low energy determinations when
extrapolated to the $Z$-pole.  The lineshape determination, however, is
sensitive to the presence of new physics which affects the
$Zb\bar{b}$ vertex or the hadronic widths.

\item The agreement between the indirect prediction for $m_t$
with the tentative direct CDF observation
and of $\alpha_s$ with the various other determinations
is an impressive success for the
entire program of precision observables.

\item Combining the direct CDF value of $m_t$ with the indirect constraints
does not make a large difference within the context of the standard model.
However, when one goes  beyond the standard model, the direct $m_t$
allows a clean extraction of the new physics contributions to $\rho_0$,
which is now shown to be very close to unity, $\rho_0 = 1.0012 (17)(17)$.
This strongly limits Higgs triplet vacuum expectation values and
non-degenerate heavy multiplets.  Similarly, it allows an extraction of the
new physics contributions to $S_{\rm new}$, $T_{\rm new}$, $U_{\rm new}$,
which are consistent with zero.  Finally, one can determine the new physics
contributions to the $b\bar{b}$ vertex: $\delta^{\rm new}_{bb}$ is
approximately $2.3\sigma$ away from zero, reflecting the large value of the
$b\bar{b}$ width.

\item The data exhibit a slight preference for a light Higgs, but this is
not very compelling statistically.  One finds only $M_H \leq 570 (880)$~GeV
at 90(95\%) CL.  Furthermore, the preference depends crucially on the large
observed value of $\Gamma(b\bar{b})$, and to a lesser extent on the
large SLD value for $A^0_{LR}$.  Omitting these
values the  $M_H$ dependence of the observables is weak.

\item The major prediction of supersymmetry is that one does not expect
large deviations in the precision observables.  The new particles tend to
be heavy and decouple.  One implication that is relevant, however, is that
supersymmetric theories have a light standard model-like Higgs.  They
therefore favor the lighter Higgs mass and the lower end of the predicted
$m_t$ range.  Also, the observed gauge couplings are consistent with the
coupling constant unification expected in supersymmetric grand unification,
but not with the simplest version of non-supersymmetric unification.

\item In compositeness and dynamical symmetry breaking theories one
typically expects not only large flavor changing neutral currents but
significant deviations of $\rho_0$ from unity and of $S_{\rm new}$ and
$T_{\rm new}$ from zero.  One further expects that $\delta_{bb}^{\rm new} <
0$, at least in the simplest models.  Therefore, the precision experiments
are a major difficulty for this class of models.

\end{itemize}

\vglue 0.2cm

\end{document}